\documentclass[fleqn,usenatbib]{mnras}
\usepackage{newtxtext,newtxmath}

\usepackage[T1]{fontenc}

\DeclareRobustCommand{\VAN}[3]{#2}
\let\VANthebibliography\thebibliography
\def\thebibliography{\DeclareRobustCommand{\VAN}[3]{##3}\VANthebibliography}

\usepackage{graphicx}	
\usepackage{amsmath}	
\usepackage{threeparttable}
\usepackage{xcolor}

\title[Parsec-scale properties of 0858$-$279]{Parsec-scale properties of the peculiar gigahertz-peaked spectrum quasar 0858$-$279}

\author[Kosogorov et al.]{\parbox{\textwidth}{
N. A. Kosogorov,$^{1,2}$\thanks{E-mail: nakosogorov@gmail.com}
Y. Y. Kovalev,$^{2,1,3}$
M. Perucho,$^{4,5}$
Yu. A. Kovalev$^{2}$
}
\vspace{0.4cm}\\
\parbox{\textwidth}{
$^{1}$Moscow Institute of Physics and Technology, Institutsky per.~9, Dolgoprudny 141700, Russia\\
$^{2}$Lebedev Physical Institute of the Russian Academy of Sciences,
Leninsky prospekt 53, 119991 Moscow, Russia\\
$^{3}$Max-Planck-Institut für Radioastronomie, Auf dem H\"ugel 69, D-53121 Bonn, Germany\\
$^{4}$Departament d'Astronomia i Astrof\'{\i}sica, Universitat de Val\`encia, C/ Dr. Moliner, 50, E-46100, Burjassot, Val\`encia, Spain\\
$^{5}$Observatori Astron\`omic, Universitat de Val\`encia, C/ Catedr\`atic Jos\'e Beltr\'an 2, E-46980, Paterna, Val\`encia, Spain
}}

\date{Accepted 2021 December 4. Received 2021 December 3; in original form 2021 April 17}

\pubyear{2021}

\begin{document}
\label{firstpage}
\pagerange{\pageref{firstpage}--\pageref{lastpage}}
\maketitle

\begin{abstract}
We performed multi-frequency studies on the gigahertz-peaked spectrum high-redshift quasar 0858$-$279. Initially, the source presented itself in the early VLBI images as a very peculiar resolved blob. We observed the quasar with the VLBA at 1.4$-$24~GHz in a dual-polarization mode. The high spatial resolution and the spectral index maps enabled us to resolve the core-jet structure and locate a weak and compact core by its inverted spectrum. The dominant jet component 20 parsecs away from the core was optically thin above 10~GHz and opaque below it. We also estimated an uncommonly strong magnetic field in the bright jet feature, which turned out to be around 1~G. The Faraday rotation measure maps revealed high RM values over 6000~rad/m$^{2}$. Additionally, these maps allowed us to follow the magnetic field direction in the bright jet feature being perpendicular to the propagation direction of the jet. All the results strongly indicated the formation of a shock wave in the dominant component arising from an interaction with the surrounding matter. Using the proposed hypothesis and the core shift approach, we discovered that the magnetic field in the core region is of the order of 0.1~G. 
\end{abstract}



\section{Introduction}

Gigahertz-peaked spectrum (GPS) active galactic nuclei (AGN) are powerful radio sources which usually interact with the dense gas surrounding them \citep[e.g.][]{1998PASP..110..493O,2020arXiv200902750O}. Supposedly, they are going through the early stages of evolution. Therefore, it is of great interest to investigate such sources to learn about the AGN formation and its interaction with nearby matter. Several studies have been conducted to determine the nature of GPS sources \citep[e.g.][]{2001A&A...377..377S,2003PASA...20..151J,2005A&A...435..839T,2009AN....330..199S,2013AstBu..68..262M}. The turnover in their spectra is often associated with the synchrotron self-absorption of relativistic electrons. Some common properties of GPS sources are the lack of strong flux density variability, low speed of jet components and low degree of linear polarization. However, there are many exceptions since the definition of GPS objects is solely based on their spectra. At the same time, GPS quasars can have very complex and different physics behind them. The discussion of classification problems is presented in, e.g. \citet{2003ASPC..300...71L}. For instance, there are ``masquerading'' blazars with spectra which can change drastically with time \citep[e.g.][]{2002PASA...19...83K,2005BaltA..14..413K} and sometimes may show convex profiles due to the ejection of new material from the core. A small number of sources showed that shocks in their jets could be formed as a result of interaction with the ambient medium \citep[e.g.][]{2014SSRv..183..405H}. As a consequence, they can have a regular GPS-like spectrum.

\begin{figure*}
    \includegraphics[width=0.39\linewidth,trim=0cm 0cm 0cm 0cm,clip]{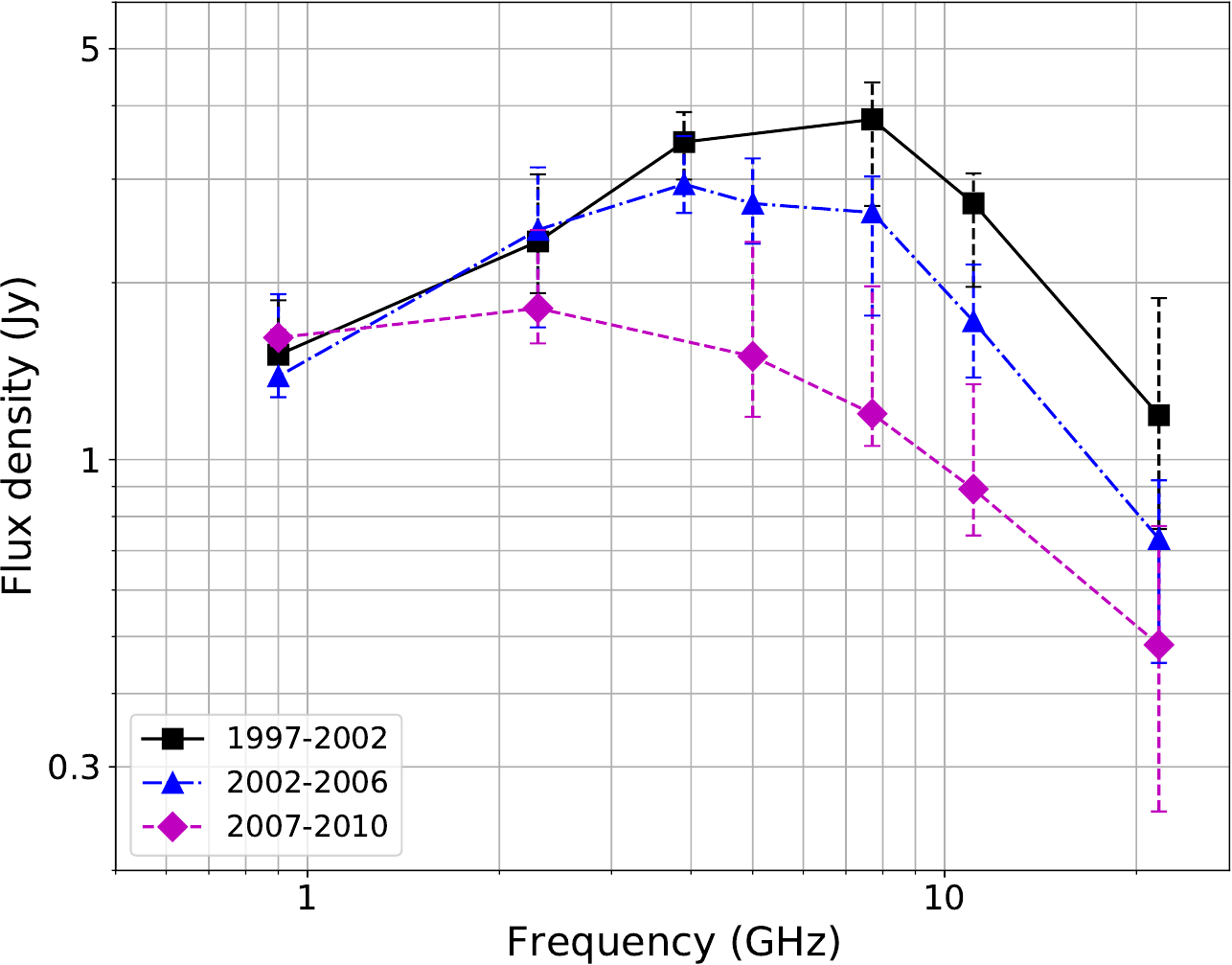}%
    \quad
    \includegraphics[width=0.56\linewidth,trim=0cm 0cm 0cm 0cm,clip]{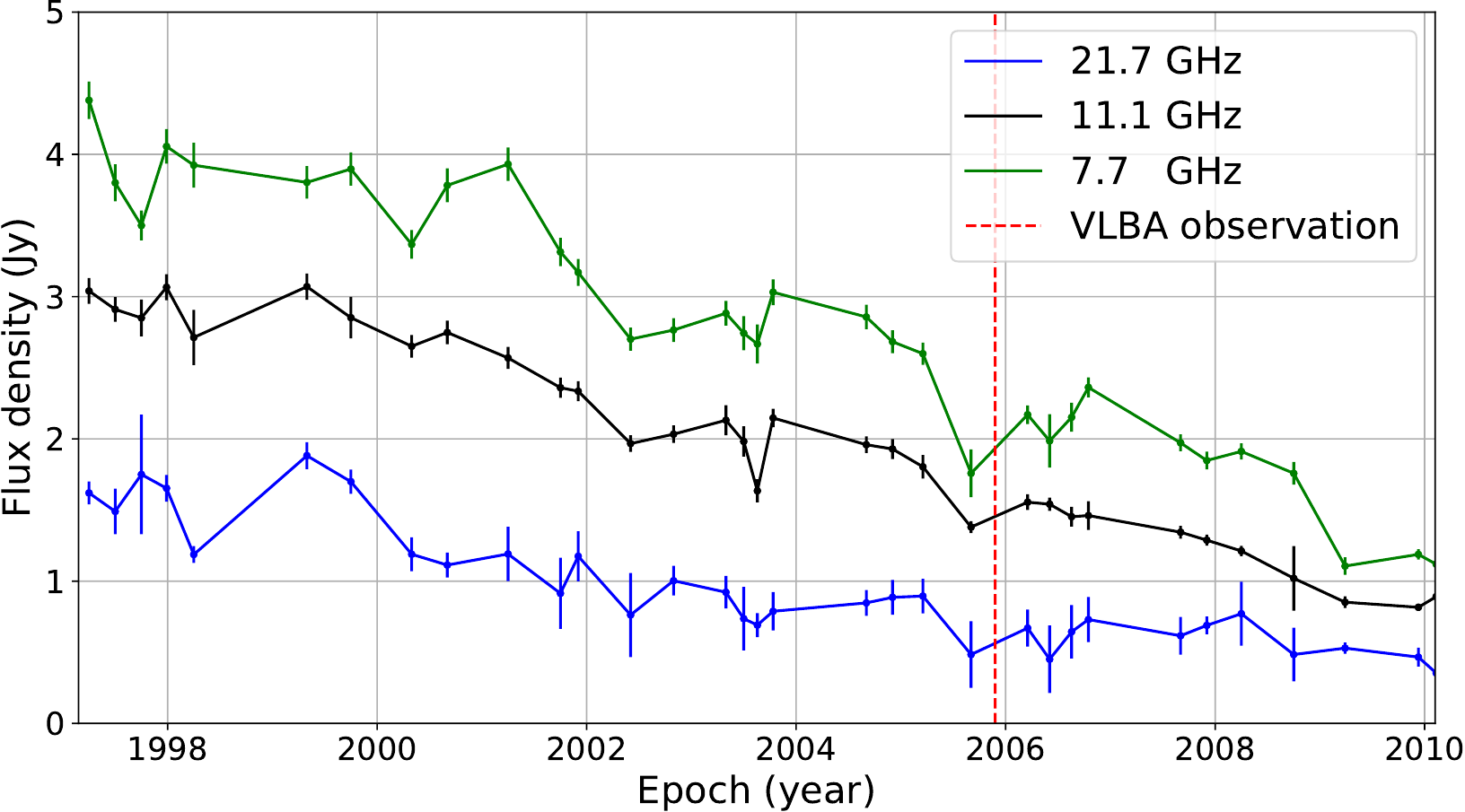}
    \caption{Median RATAN-600 spectra over three indicated time periods of observations for the GPS quasar 0858$-$279 (left) and multi-frequency RATAN-600 light curve of the source for three frequencies (right). The dashed vertical lines in the left picture show the spread of values from the lowest to the highest of flux density for a given interval of years. The red dashed line in the right picture shows the time of the VLBA observation.
    The flux density $S$ changes by more than a factor of 2 over the variability time, whose scale is of the order of several months. }
    \label{fig:RATAN_spectra_flux_fig}
\end{figure*}


Very Long Baseline interferometry (VLBI) observations provide parsec-scale radio images of distant sources. It is a unique tool to image sources with unprecedented high angular resolution. Usually, observed AGN have a core-jet structure, where the VLBI core is typically the brightest feature located at the visible beginning of the jet. Ultra-relativistic electrons are the main source of synchrotron emission in AGN. The apparent core is associated with an opaque region with a flat spectrum where the optical depth reaches $\tau_\mathrm{depth} \approx 1$ \citep{1979ApJ...232...34B}, while the jet typically shows a steep spectrum of optically thin synchrotron radiation. 
Throughout the paper, the sign of the spectral index $\alpha$ is defined as $S \propto \nu^{+\alpha}$, where $S$ is the flux density observed at frequency $\nu$. Many works have been dedicated to study the spectral index properties \citep[e.g.][]{1979ApJ...231..299R,2011A&A...535A..24S,2012A&A...544A..34P,2014AJ....147..143H,2019MNRAS.485.1822P}. For most jets, the spectral index $\alpha$ has negative values.

In this paper, we study the GPS quasar PKS\,0858$-$279, also known as PMN J0900$-$2808. Previous observations of this quasar revealed surprising properties. Monitoring at RATAN-600 \citep{1999A&AS..139..545K,2002PASA...19...83K} of compact extragalactic radio sources showed that the object has a convex gigahertz-peaked continuum radio spectrum (\autoref{fig:RATAN_spectra_flux_fig}). We note that the peak frequency at which the flux density maximum is reached drops from 5.5 to 3.3 GHz over the period of ten years of observations. Apart from that, the VLBA Calibrator Search program data \citep{2002ApJS..141...13B} were used to analyse the milliarcsecond scale structure of the quasar. The VLBI images demonstrated that the quasar looks like a typical extended GPS source at centimetre wavelengths. However, it is distant, with redshift $z = 2.152$~\citep{1993A&AS...97..483S} corresponding to young universe, only 3 Gyr since the Big Bang, and it exhibits a high amplitude of radio variability (\autoref{fig:RATAN_spectra_flux_fig}), which is also supported by the ATCA monitoring results \citep{2003PASJ...55..351T,2004A&A...424...91E}. Additionally, the dominating component was found to be too large (1.3 mas or about 10~pc at 4~cm and 12~mas or about 95~pc at 13 cm, projected) to account for the observed timescale of variability, which is of the order of several months. The corresponding light travel time does not match the observed size. This object, as reported by \citet{2004A&A...415..549R}, has a high degree of linear polarization. It is quite uncommon for GPS radio sources. In this paper, we investigate what stands behind such a peculiar parsec-scale structure and the behaviour of the spectrum.

This paper is structured as follows: in \autoref{sec:data}, the observations and the data reduction are described. In \autoref{sec:doppler}, we derive the variability Doppler factor from the multi-frequency light curve.  In \autoref{sec:Imaps}, we present the total intensity maps and the integrated VLBA spectrum. Aside from that, we analyse the spectral index maps and the structure of the quasar. In \autoref{sec:magnetic}, we estimate the magnetic field in the dominating component. \autoref{sec:polar} is dedicated to polarization, Faraday rotation measurements and the spatial structure of the magnetic field. In \autoref{sec:origin}, a possible explanation of the bright detail origin and the core magnetic field estimation are discussed. Finally, \autoref{sec:summary} summarises all the results. Through the paper, the model of flat $\mathrm{\Lambda C D M}$ cosmology was used with $H_0 = 68\,\mathrm{km\,s}^{-1}$, $\Omega_M$ = 0.3 and $\Omega_\Lambda=0.7$~\citep{2016A&A...594A..13P}. The position angles are measured from north through east.

\begin{figure*}
	\includegraphics[width=\linewidth,trim={4cm 18.6cm 4cm 4cm},clip]{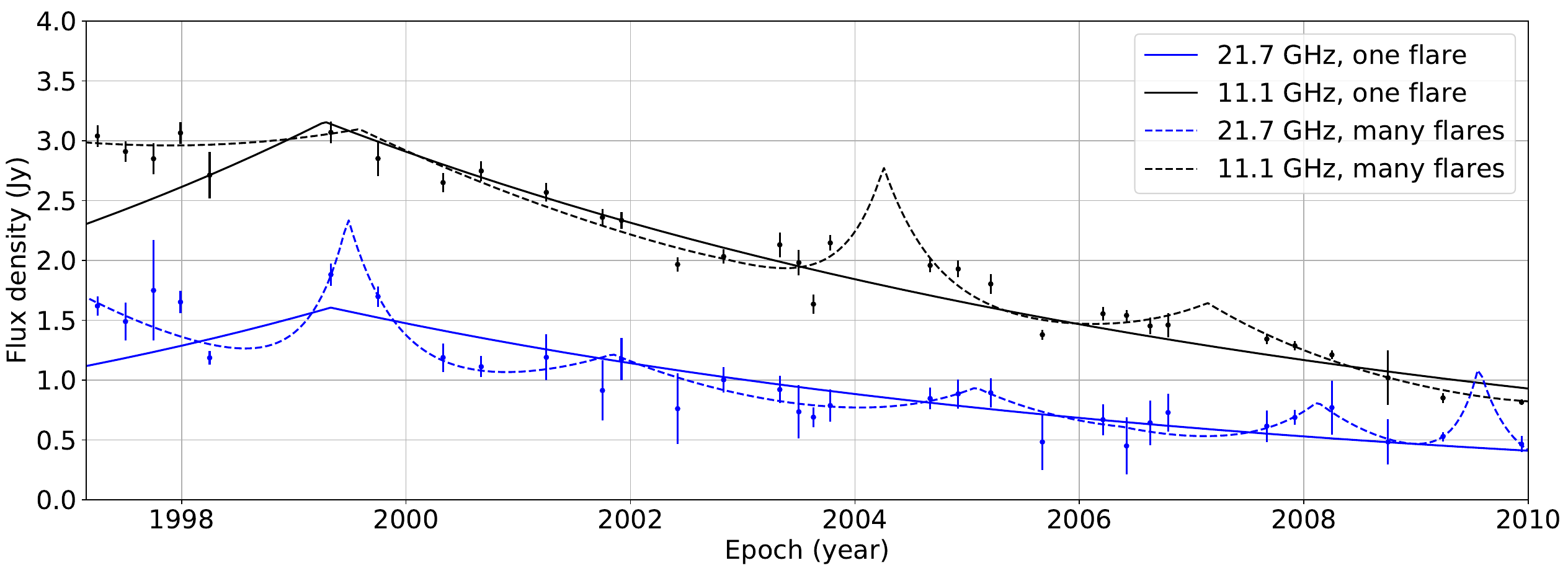}
    \caption{RATAN-600 flux density light curves decomposed into one (solid curves) and many flares (dashed curves) for 21.7~GHz and 11.1~GHz. The solid curves show the conservative single-flare model which is used to derive the lower limit on the Doppler factor. }
    \label{fig:flux_time_model_fig}
\end{figure*}

\section{Observations and Data reduction}
\label{sec:data}
\subsection{RATAN-600 observations}

The object has been observed at 5-7 frequencies from 1 to 22~GHz with the RATAN-600 radio telescope of the Special Astrophysical Observatory \citep[e.g.][]{1993IAPM...35....7P} since 1997 within 2-4 sets per year. In this article we show and use in the subsequent analysis the data from the 1997-2010 period.
The source is a member of the monitoring sample of about 600 VLBI-selected extragalactic objects.
Typically, continuum spectra for every source were measured 1-3 times in each observing set and averaged. Each spectrum was observed within a few minutes in a transit mode.
The amplitude calibration curve at every frequency was independently constructed for every observing set from measurements of up to ten calibrators mostly selected from \cite{1977A&A....61...99B,1994A&A...284..331O}. The flux density is calibrated according to \cite{1994A&A...284..331O}.
All measurement errors were calculated as weighted averages, and the measurement errors also included amplitude calibration curve uncertainty. The flux density scale error is not included. 
Increased error values seen in \autoref{fig:RATAN_spectra_flux_fig} and \ref{fig:flux_time_model_fig} typically reflect measurements made in non-optimal weather conditions and, in some cases, radio frequency interference.
More details of the measurement procedure can be found in \citet{1999A&AS..139..545K}.

\subsection{VLBA observations}

The observation of the quasar was performed on 26~November~2005. The project code is BK128. It was made in a dual-polarization mode with the VLBA at observing bands centred at 1.5~GHz, 2.3~GHz, 4.8~GHz, 8.3~GHz, 15.4~GHz and 22.2~GHz. Each band consisted of four 8~MHz-wide frequency channels (IFs) per polarization. For the further analysis of Stokes I maps, spectral properties and the estimation of the magnetic field, we combined all four IFs at all frequencies. For the polarization analysis, the four lowest frequency bands were split into two sub-bands, two IFs per sub-band, centred at 1.41~GHz, 1.66~GHz, 2.28~GHz, 2.39~GHz, 4.64~GHz, 5.04~GHz, 8.15~GHz, 8.50~GHz. The total observing time at four lowest frequencies was about 50~min per frequency. The total time at 15.4~GHz and 22.2~GHz was 60 and 65~min, respectively. 
The data at each VLBA station was recorded with 2-bit sampling and a total bit rate of 256~Mbps.

In addition to the investigated quasar 0858$-$279, various calibrators were observed. They included 
fringe check, cross-polarization and band-pass calibrator
1226$+$023 (3C\,273), electric vector position angle (EVPA) calibrators 0851$+$202 (OJ\,287) and 1328+307 (3C\,286) and the leakage-term calibrator 0919$-$260.

The initial data processing was carried out in the Astronomical Image Processing System package~\citep{2003ASSL..285..109G} for each sub-band independently and included typical VLBI data processing procedures described in detail in the AIPS cookbook. As the initial phase self-calibration step of imaging, the CALIB task in AIPS was used. Hereafter, CLEAN imaging, phase and amplitude self-calibration were iteratively done in the difmap software package \citep{1997ASPC..125...77S}. The resulting amplitude calibration accuracy was estimated to be about 10\,\% for 22~GHz and about 5\,\% for frequencies in the 1.4~GHz to 15~GHz range. These estimations are consistent with the calibration accuracy described in the VLBA observational status summary\footnote{\url{https://science.nrao.edu/facilities/vlba/docs/manuals}}. 

To calibrate the EVPA (see \autoref{sec:polar}) of the investigated quasar, we used 3C\,286 for 1.5~GHz and 2.2~GHz bands and OJ\,287 for all other higher frequencies. The absolute EVPA values for 3C\,286 were taken from the guide to observing with the VLA\footnote{\url{https://science.nrao.edu/facilities/vla/docs/manuals/obsguide}}. The position angle of this source is known to be constant over time and equal to 33$^\circ$ in a wide range of low frequencies from 1 to 7 GHz \citep{2013ApJS..206...16P}. Apart from that, VLBI images for this source showed that the parsec-scale polarized emission could be consistently calibrated with the VLA value \citep[e.g.][]{1996A&A...312..380J,1997A&A...325..479C}.
The absolute EVPA values for OJ\,287 were taken from the VLA (Very Large Array) Monitoring Program\footnote{\url{http://www.vla.nrao.edu/astro/calib/polar/}} \citep{taylor2000vlba} and the MOJAVE\footnote{\url{http://www.physics.purdue.edu/astro/MOJAVE/}} (Monitoring Of Jets in Active galactic nuclei with VLBA experiments) observations.
We note that OJ\,287 is variable; it was observed with the VLA at the day of our VLBA observations (26 November 2005), which has allowed for an accurate absolute EVPA calibration.
We have taken into account that the frequencies in our observations differ from the VLA and MOJAVE ones. For that, we fit the linear model to EVPA -- $\lambda^2$ VLA data for OJ\,287, solving the $n\pi$ ambiguity problem (see \autoref{sec:polar} for more detail). We used that model to determine the EVPA values at the frequencies of the VLBA observations. For all bands, the EVPA accuracy of the VLA observations reported on the corresponding website was smaller than 3$^\circ$. 
The MOJAVE OJ\,287 EVPA value agreed well with the fitted model within its error. However, we did not use MOJAVE data in the fit since MOJAVE observing epoch was more than hundred days away from our VLBA date.
The error of absolute EVPA calibration was calculated from the linear regression fit. The details on how the total EVPA error was calculated are presented in \autoref{sec:polar}.


\begin{figure*}
	\begin{center}
    \includegraphics[width=0.45\linewidth, trim=1cm 1.5cm 1cm 1.5cm,clip]{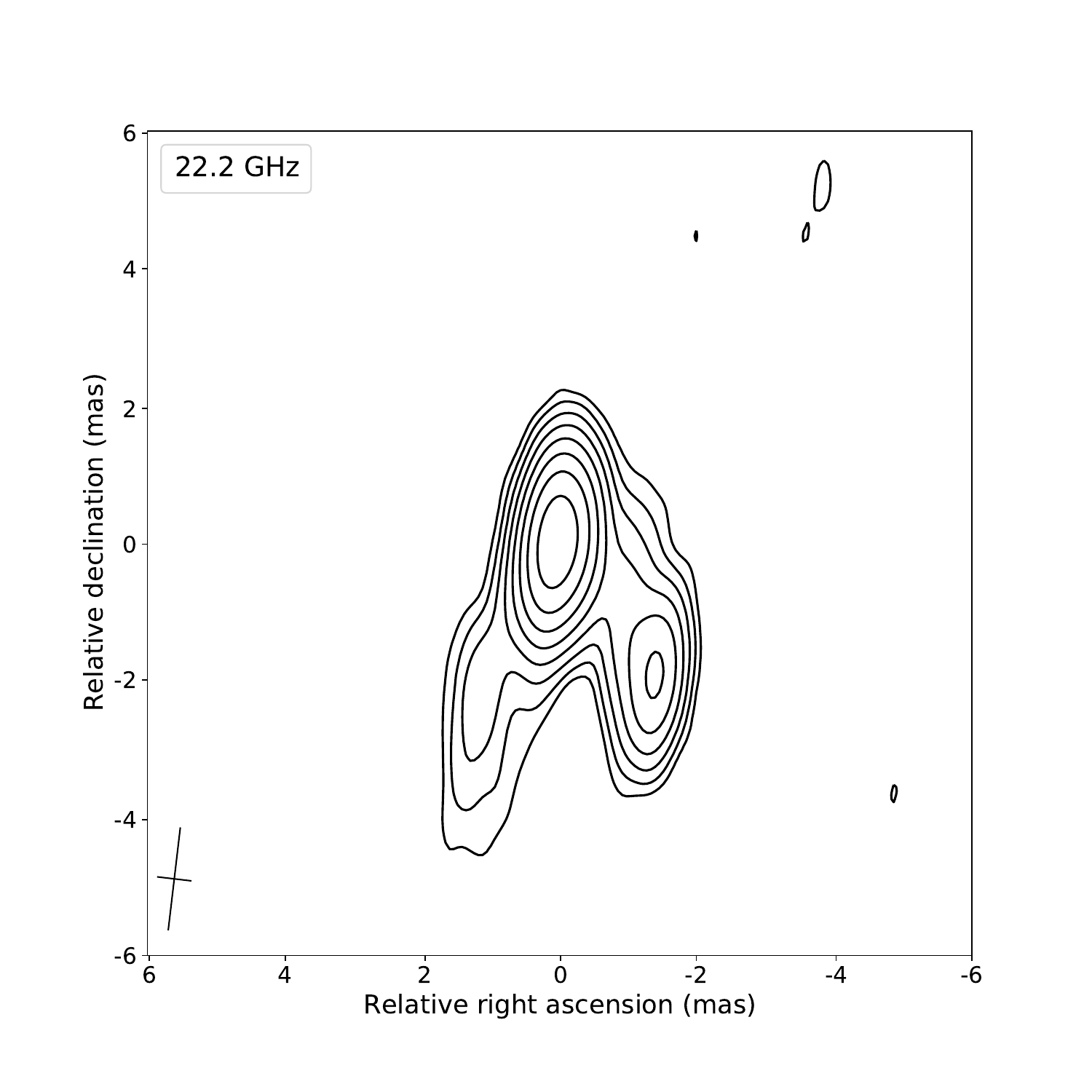}%
    \includegraphics[width=0.45\linewidth,trim=1cm 1.5cm 1cm 1.5cm,clip]{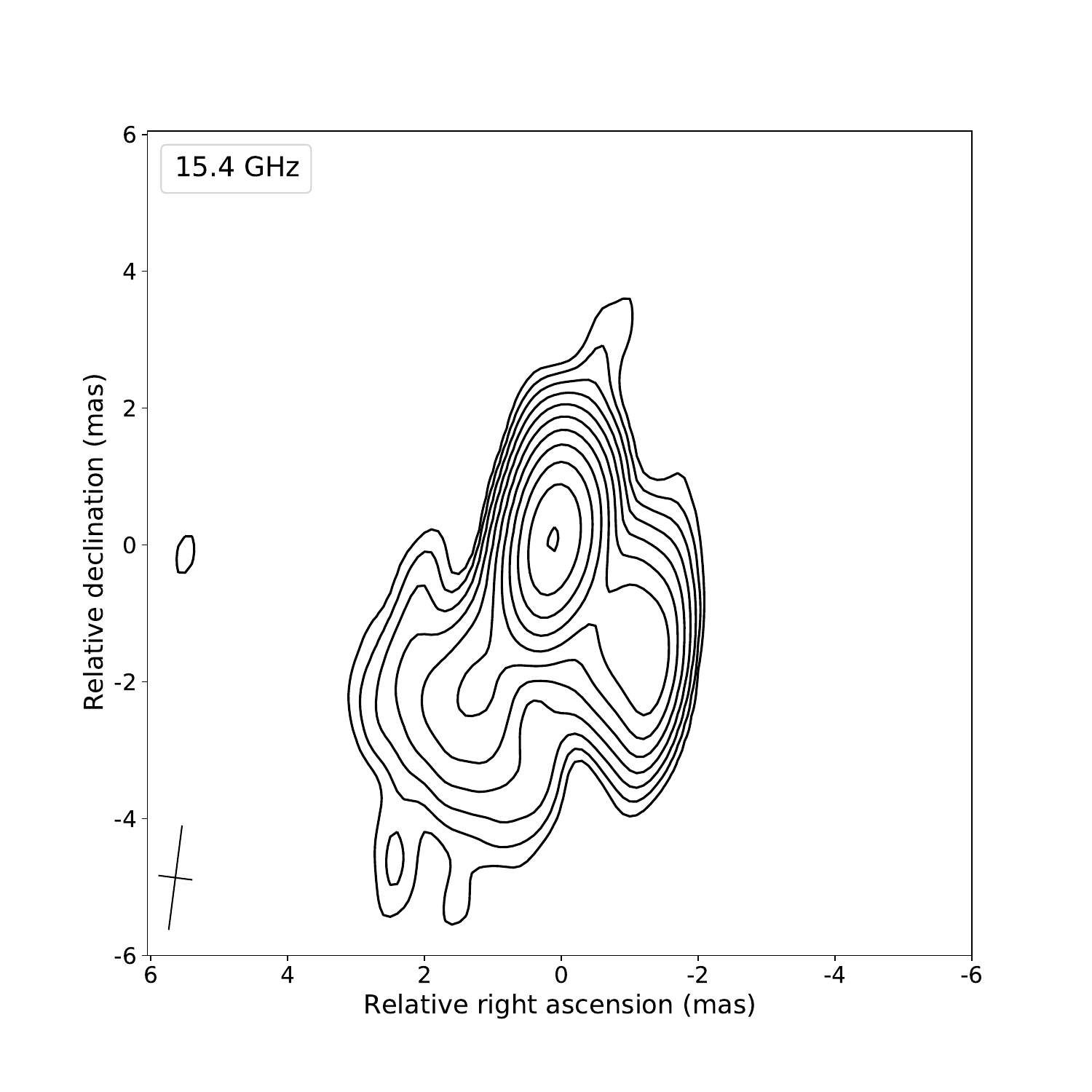}

    \includegraphics[width=0.45\linewidth, trim=1cm 1.5cm 1cm 1.5cm,clip]{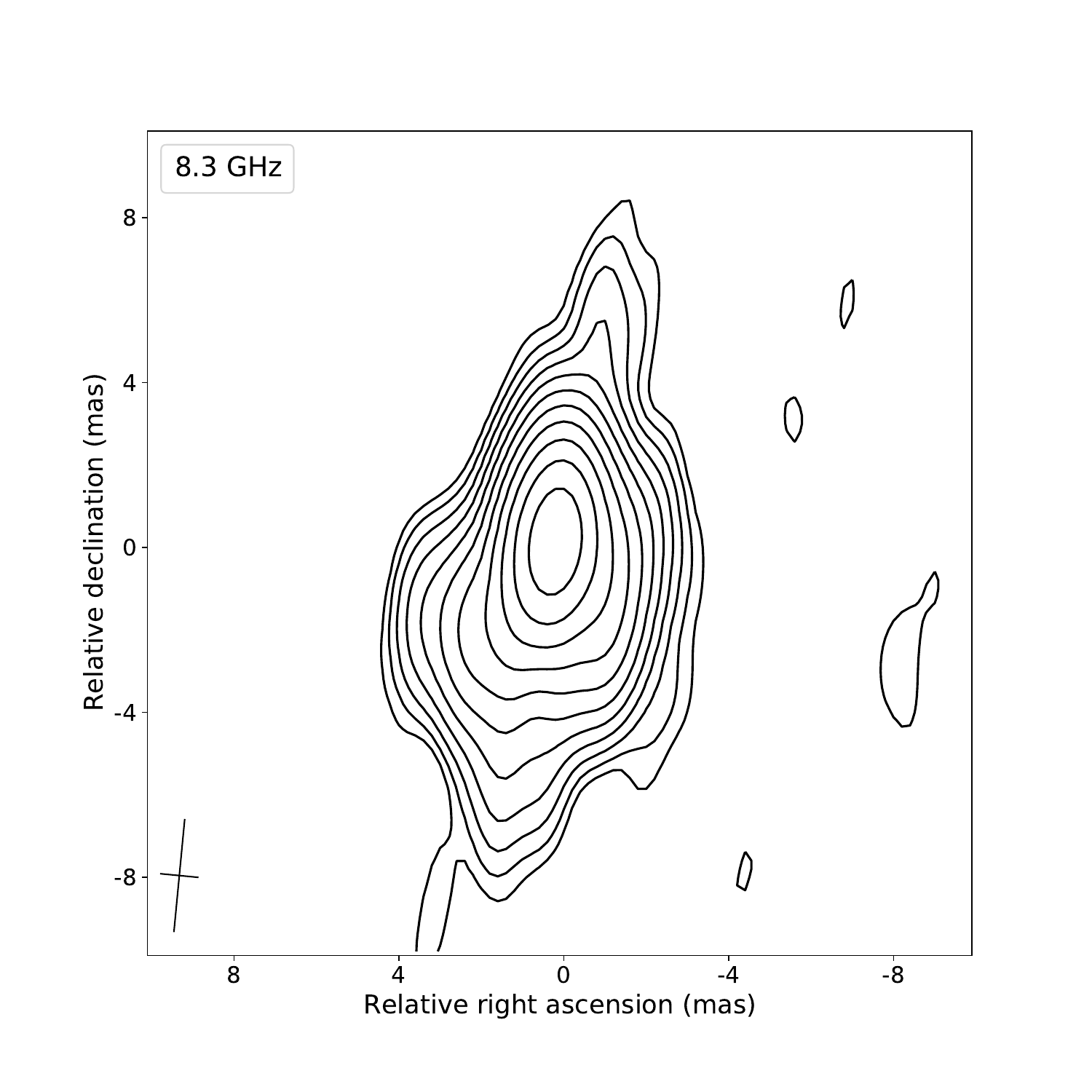}%
    \includegraphics[width=0.45\linewidth,trim=1cm 1.5cm 1cm 1.5cm,clip]{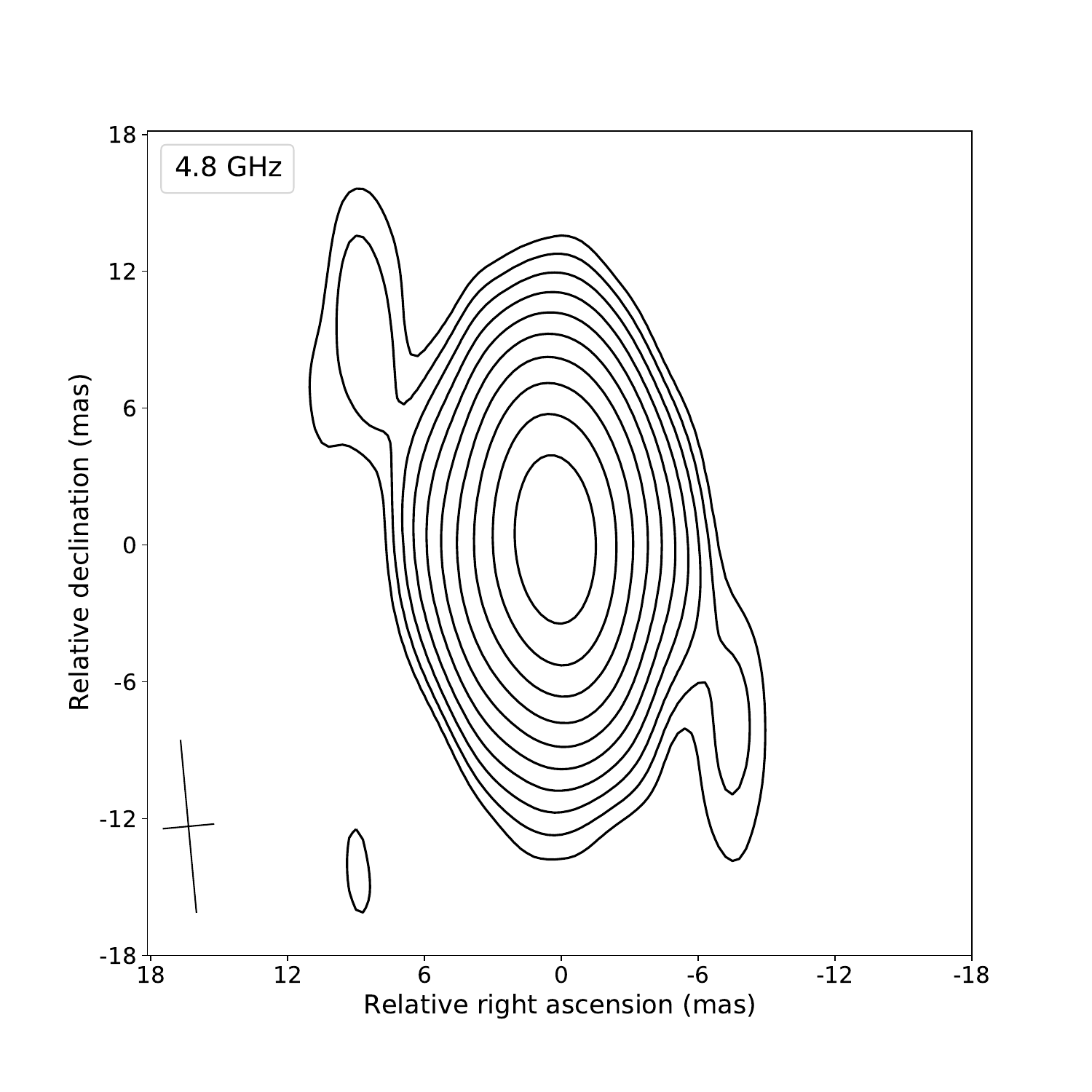}

    \includegraphics[width=0.45\linewidth, trim=1cm 1.5cm 1cm 1.5cm,clip]{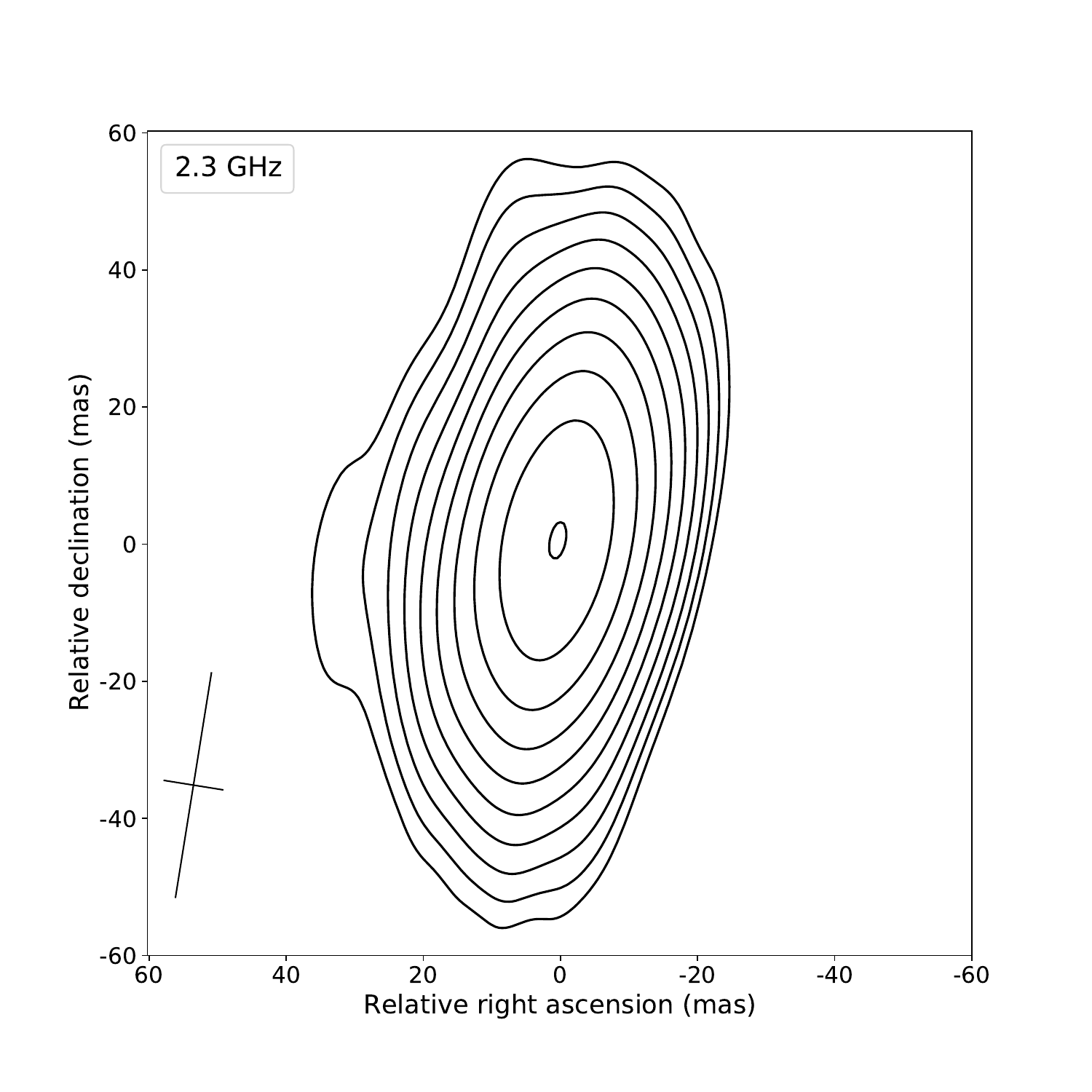}%
    \includegraphics[width=0.45\linewidth,trim=1cm 1.5cm 1cm 1.5cm,clip]{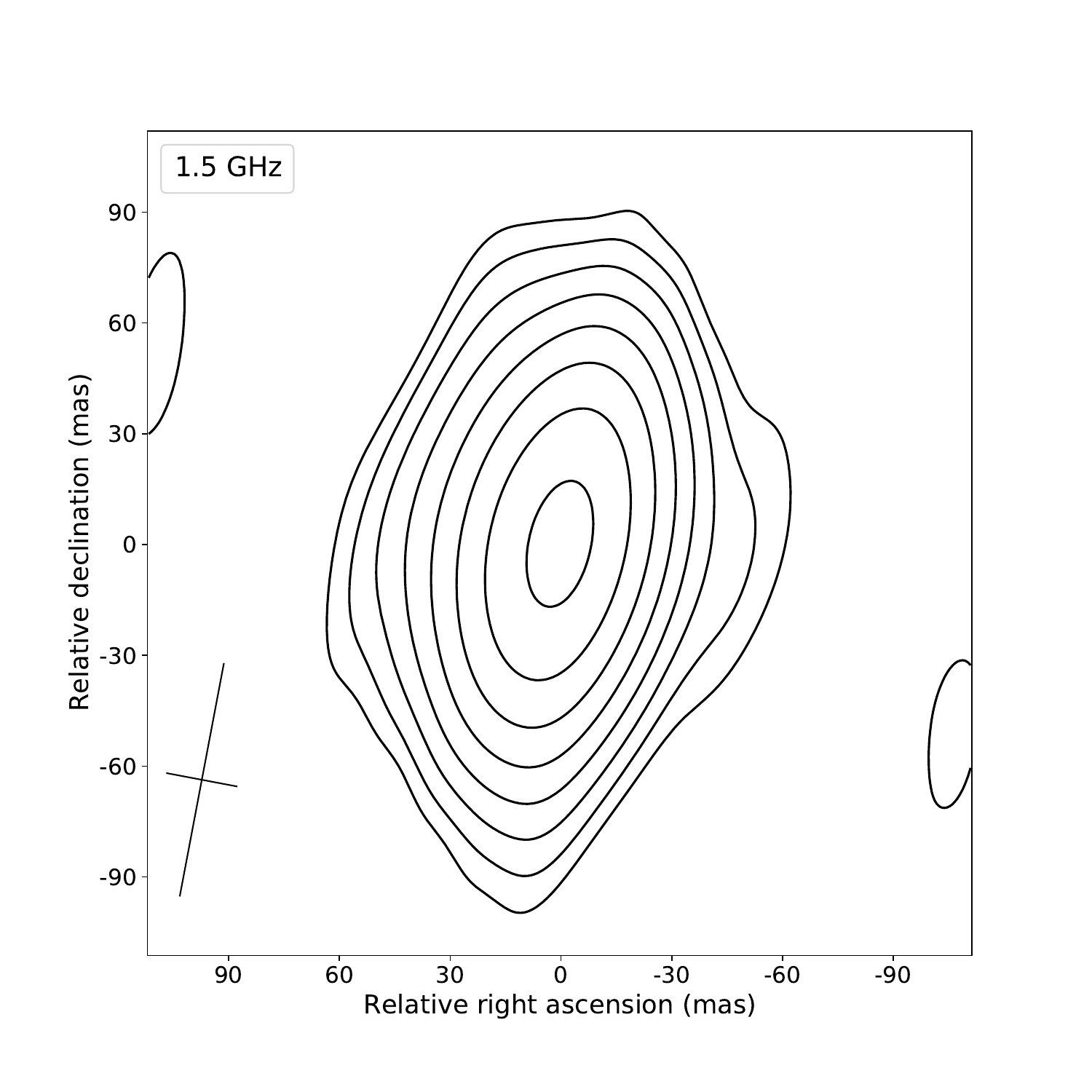}
    \end{center}
    \caption{Total intensity multi-frequency contour maps for the quasar 0858$-$279 from the epoch 2005-11-26. The contours start at the 2$\sigma_I$ base level and are plotted in $\times2$~steps. The beam is shown at the FWHM level in the left bottom corner. The basic parameters of the maps are presented in \autoref{tab:intensity_table}.}
    \label{fig:intensity_fig}
\end{figure*}

\section{Variability Doppler factor}
\label{sec:doppler}

There are several approaches suggested to estimate the Doppler boosting factors, such as analysing X-ray observations with VLBI fluxes \citep[e.g.][]{1993ApJ...407...65G} or directly observing the brightness temperature with the help of VLBI. Apart from that, one can compare the flux decline time of a component in the jet with its VLBI size \citep{2005AJ....130.1418J} or find the Doppler factors from the total flux density observations over time, as we used in this paper \citep{1999ApJ...511..112L}. To obtain the variability Doppler factor of the jet, we need to decompose each flux curve into a number of flares. Afterwards, the parameters of each flare can be calculated with the aim to obtain the observed brightness temperature of the source. The observed brightness temperature is boosted compared to the intrinsic one $T_\mathrm{int}$. Although in some papers \citep[e.g.][]{2009A&A...494..527H}, it is assumed to be close to the equipartition value $T_\mathrm{eq} = 5 \times 10^{10}$~K. We believe that for flares, the value is closer to the inverse-Compton catastrophe limit \citep{1994ApJ...426...51R}. Thus, for the intrinsic temperature, we use the value $T_\mathrm{int} = 2.78 \times 10^{11}$~K obtained in \citet{2018ApJ...866..137L}.

The flares were modelled using the exponential form which follows  \citet{1999ApJS..120...95V, 2009A&A...494..527H}:
\begin{equation}
    \Delta S(t)=\left\{
                \begin{array}{ll}
                  \Delta S_\mathrm{max} e^{(t-t_\mathrm{max})/\tau},\;\;\;\;\;\;\;\; t<t_\mathrm{max}\\
                  \Delta S_\mathrm{max} e^{-(t-t_\mathrm{max})/1.3\tau},\;\; t>t_\mathrm{max}
                \end{array}
              \right.,
	\label{eq:flares_eq}
\end{equation}
where $\Delta S_\mathrm{max}$ is the maximum flux density of the flare in Jy observed at the time $t_\mathrm{max}$ and $\tau$ is the rise time of the flare. We followed \citet{1999ApJS..120...95V} taking the coefficient 1.3 in the decay timescale ratio for the sake of consistency.

We used two RATAN-600 light curves at 21.7~GHz and 11.1~GHz to fit the flares because the variability timescales at high frequencies are typically smaller. As a consequence, the problem of blending of flares is less severe. We used the one-flare model, considering that all data points can be described as a single prolonged flare. This set a lower limit on the Doppler factor. 
Apart from that, we decomposed our light curves into several flares. Since the number of the data points is not large in our observations, we applied the Akaike information criterion \citep[e.g.][]{1974ITAC...19..716A} to find the suitable number of flares representing our curves. All the model curves are shown in \autoref{fig:flux_time_model_fig}.

Using the obtained fits, we calculated the associated
variability brightness temperature in K described in \citet{1979ApJ...232...34B} of each flare as:
\begin{equation}
    T_\mathrm{var} = 1.47 \times 10^{13} \frac{D_L^{2} \Delta S_\mathrm{max}(\nu)}{\nu^{2}\tau^{2}(1+z)^{4}} \,,
	\label{eq:br_temp_eq}
\end{equation}
where $D_L$ is the luminosity distance in Mpc, $\tau$ is
the variability time-scale in days, $z = 2.152$ is the redshift of the quasar, $\nu$ is the observed frequency in GHz. The corresponding variability Doppler factor of a flare, following eq. (2) in \citet{2017MNRAS.466.4625L}, is calculated as:
\begin{equation}
    \delta_\mathrm{var} = (1+z) \left( \frac{T_\mathrm{var}}{T_\mathrm{int}} \right)^{1/3}\,.
	\label{eq:doppler_eq}
\end{equation}

Ultimately, using a one-flare model for 11 and 22~GHz, the obtained averaged Doppler factor turned out to be $\delta_\mathrm{var}=2.1 \pm 0.1$. The error was calculated from the variance–covariance matrix of the fit. Using many flares for the two frequencies, we managed to estimate the variability Doppler factor of the source as the median value of all the obtained factors for each peak $\delta_\mathrm{var} = 4.8 \pm 3.9$. Both approaches give the same order of magnitude. However, the model with several peaks may be untrustworthy; we prefer to use the conservative approach with the single prolonged flare and the estimated lower limit in the subsequent analysis.

\section{Multi-frequency total intensity maps}
\label{sec:Imaps}

\subsection{Stokes I maps and integrated VLBA spectrum}
\label{sec:Total intensity maps}

\begin{table*}
	\centering
	\caption{Main parameters of the Stokes $I$ VLBA maps from the epoch 2005-11-26.}
	\label{tab:intensity_table}
	\begin{tabular}{rrrcccr}
		\hline
		Frequency & Total flux density, $S$ &Map $I$ peak & Map noise, $\sigma_I$ & $B_\mathrm{maj}$ & $B_\mathrm{min}$ & $B_\mathrm{PA}$\\ 
		(GHz) & (mJy) & (mJy/beam) & (mJy/beam) & (mas) & (mas) & (deg) \\
		\hline
		22.2 & 552 & 358  &  0.8 & 1.5 & 0.5 & $-$6.7 \\   
        15.4 & 969 & 478  &  0.2 & 1.5 & 0.5 & $-$7.3 \\
        8.3 & 1577 & 758  &  0.3 & 2.7 & 0.9 & $-$5.5 \\
        4.8 & 2114 & 1056 &  0.6 & 7.5 & 2.2 & 5.3    \\  
        2.3 & 2062 & 1111 &  1.1 & 33  & 8.6 & $-$9.1 \\
        1.5 & 1591 & 912  &  2.9 & 64  & 19  & $-$10.7\\ 
		\hline
	\end{tabular}
	\begin{tablenotes}
            \item Total flux density $S$ was calculated as a sum of CLEAN components produced in difmap. $B_\mathrm{maj}$ and $B_\mathrm{min}$ are the major and minor axis full width at half maximum (FWHM) of the restoring beam, respectively. $B_\mathrm{PA}$ is the position angle of the beam. 
        \end{tablenotes}
\end{table*}

In this section, we present the total intensity maps for all frequencies for the investigated quasar (\autoref{fig:intensity_fig}). \autoref{tab:intensity_table} shows the basic parameters of each map. The root mean square (rms) noise $\sigma_I$ of the total intensity maps is taken and averaged over the regions which do not include the quasar structure. The restoring beam parameters for two highest frequencies are comparable due to the absence of the antenna data in Brewster at 22.2~GHz.

\begin{figure}
	\includegraphics[width=\linewidth,trim={1cm 0.5cm 2.5cm 2cm},clip]{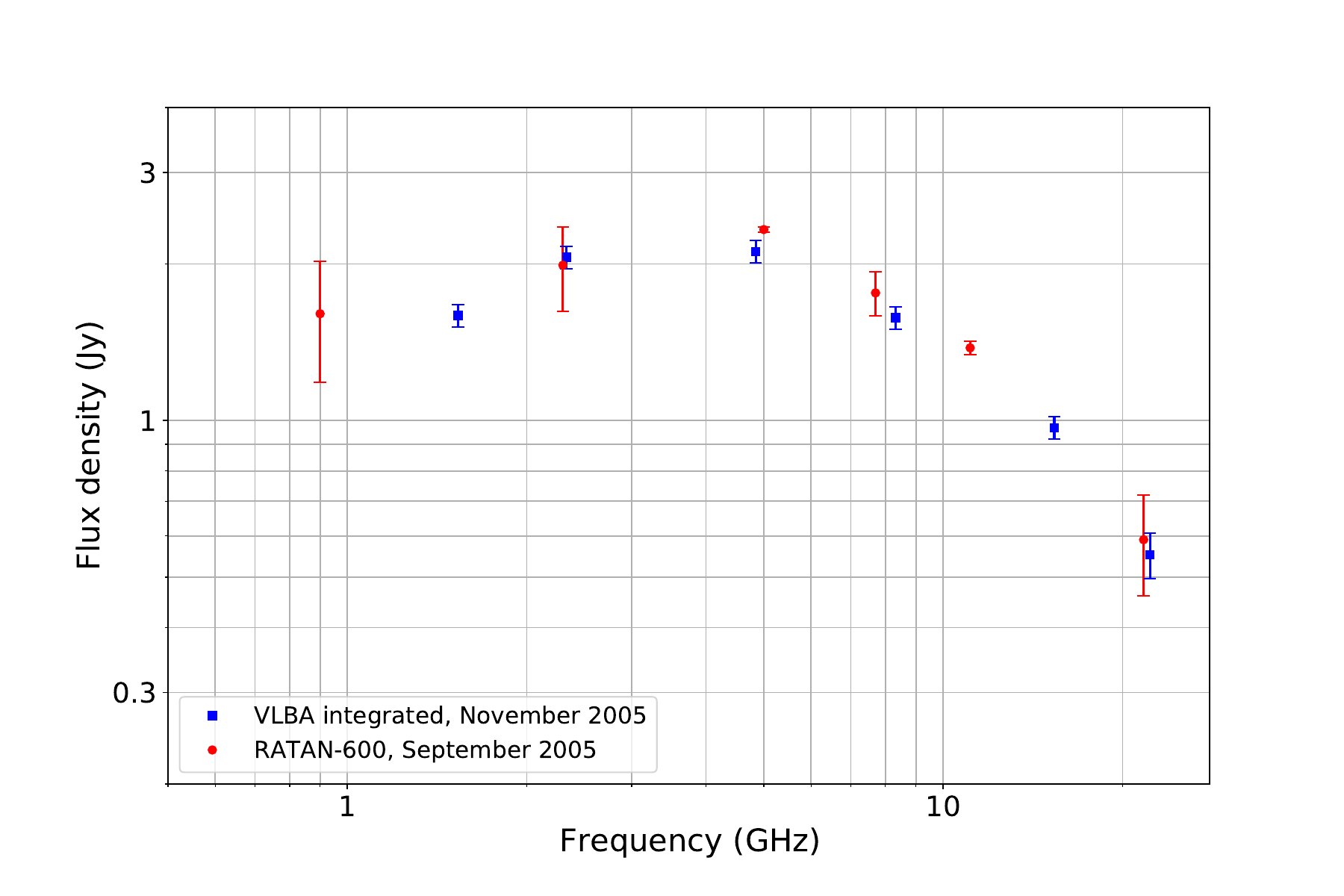}
    \caption{VLBA spectrum of the quasar (in blue) obtained from the Stokes $I$ maps and the RATAN-600 spectrum (in red) for the nearest date to the VLBA observation. The good agreement hints to the absence of any extended, unresolved flux density.}
    \label{fig:RATAN_VLBA_spectrum}
\end{figure}

The total intensity maps allowed us to obtain the integrated flux density over the field of view for every frequency. In~\autoref{fig:RATAN_VLBA_spectrum}, we present the VLBA integrated spectrum of the source with the RATAN-600 spectrum for the nearest date (2 September 2005) to the VLBA observation. One can see that the VLBA integrated flux density matches well the flux density obtained at RATAN-600. This is a consequence of good calibration and the fact that there is no extended emission observed from kiloparsec scales by the NRAO VLA Sky Survey\footnote{\url{https://www.cv.nrao.edu/nvss/}} (NVSS) \citep{1998AJ....115.1693C}. This is consistent with an assumption of the quasar being young.

\subsection{Spectral index maps}
\label{sec:spi_maps}

To locate the core and the jet and to find the direction of its flow, spectral index analysis can be performed. To reconstruct spectral index maps, image alignment should be done. It is not trivial since during the process of self-calibration, the absolute position of sources in the sky is lost. Typically, the brightest feature is shifted to become close to the phase centre of the image \citep[e.g.][]{2017isra.book.....T}. There are different methods to align images at different frequencies, e.g. taking as reference the positions of optically thin bright jet components model-fitting the source structure with 2D Gaussian components \citep[e.g.][]{2008A&A...483..759K,2011A&A...532A..38S,2013A&A...557A.105F}, using a 2D image correlation and alignment \citep[e.g.][]{2012A&A...545A.113P,2014AJ....147..143H,2019MNRAS.485.1822P}
or phase-referencing on a nearby compact calibrator \citep[e.g.][]{1984ApJ...276...56M,2018ARep...62..787V}. This paper uses the former of these methods because the absolute position of the optically thin components in the jet does not depend on frequency. 

It is necessary to model the quasar structure to apply this method. We used a set of circular Gaussian components obtained separately at the three highest frequencies in difmap. The details, including the parameters of those components, are described in \autoref{s:modeling} and \autoref{tab:gauss_comp}. By shifting the brightest Gaussian component to the centre of the image, it is possible to superimpose the images of the quasar reconstructed with the same beam parameters and the pixel size. The presented images were obtained using different $uv$-ranges at the corresponding pairs of frequencies. However, when we limited the data to have the same $uv$-ranges, spectral index images did not change within errors.

\begin{figure*}
\centering
	\includegraphics[width=\linewidth,trim={4cm 2cm 2cm 2.5cm},clip]{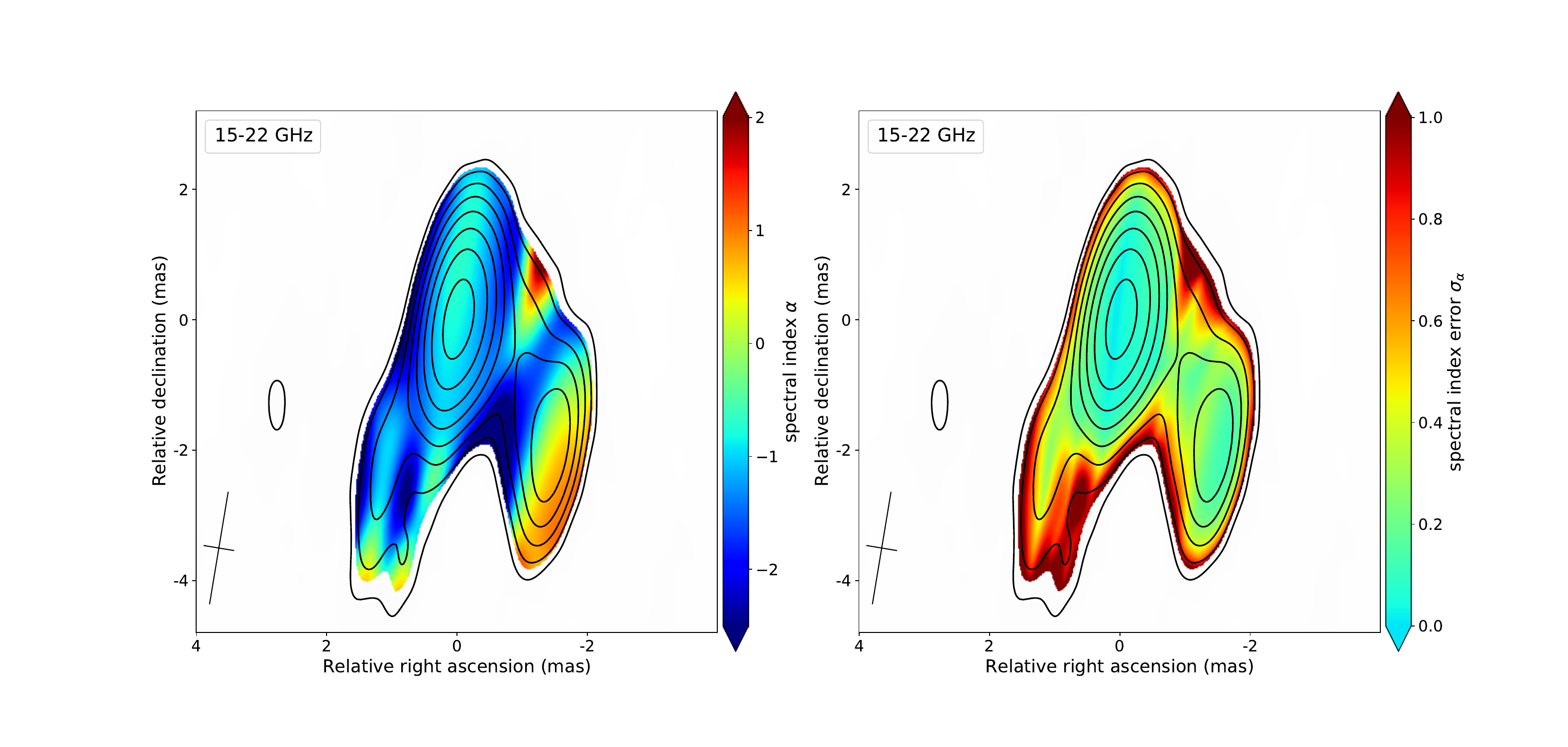}
	\includegraphics[width=\linewidth,trim={4cm 2cm 2cm 2.5cm},clip]{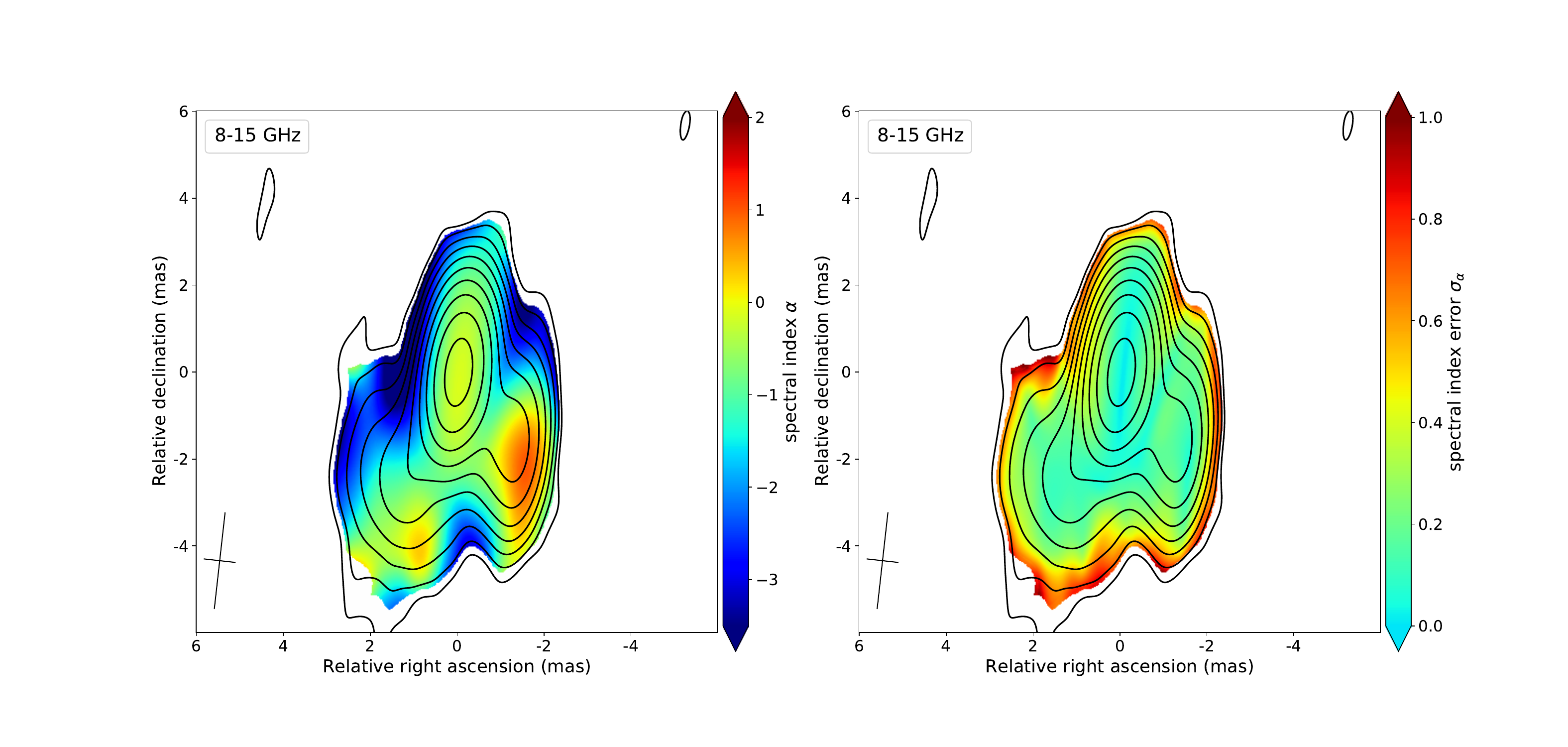}
    \caption{Spectral index (left) and spectral index error (right) maps for 15$-$22~GHz (top) and 8$-$15~GHz (bottom) frequency intervals.
    The maps show full intensity contours starting at the 2$\sigma_I$ level taken from the Stokes $I$ map for 22.2~GHz (top) and 15.4~GHz (bottom). The spectral index and the spectral index error values are shown only in those pixels where intensity exceeds 3$\sigma_I$. The beam is shown in the bottom left corner at the FWHM level.}
    \label{fig:spi_maps}
\end{figure*}

After superimposing full intensity maps on top of each other, following the alignment procedure, the spectral index was calculated for each pixel, fitting a power law to the data at two frequencies, 15.4~GHz and 22.2~GHz. The same alignment process was carried out for 8.3~GHz and 15.4 GHz. Here, we present the spectral index map and the spectral index error map for 15$-$22~GHz and 8$-$15~GHz (\autoref{fig:spi_maps}). We show only pixels in which the radiation intensity exceeds 3$\sigma_I$. The associated uncertainty was calculated using the following formula:
\begin{equation}
    \sigma_\alpha =\frac{1}{\mathrm{ln}(\nu _1/\nu _2)}\sqrt{\left ( \frac{\sigma_{I_1} }{I_1} \right )^{2}+\left ( \frac{\sigma_{I_2}}{I_2} \right )^{2}} + \sigma_{\alpha\mathrm{align}}\,,
	\label{eq:spectral_index_error}
\end{equation}
where indices 1 and 2 refer to the selected frequencies, $I$ is the radiation intensity, and $\sigma_{\alpha\mathrm{align}}$ is the error related to the alignment procedure. We followed \citet{2014AJ....147..143H} and calculated it by shifting the Gaussian component associated with the bright detail by 1~pixel (0.06~mas for 15$-$22~GHz map and 0.1~mas for 8$-$15~GHz map) in each direction. Then the standard deviation of the obtained spectral indices was calculated to estimate the alignment uncertainty.

Thus, the high-frequency VLBA observations allowed us to finally trace the true parsec-scale structure of the quasar 0858$-$279, which had not been properly traced earlier. There is a pronounced region in the spectral index image with an inverted spectrum located in the southwest, which can be identified with the quasar core marked as `C’. The emission of the dominant jet component `J' at the centre of the images is suggested to be optically thin at 15.4~GHz and 22.2~GHz since the spectral index is negative in that region. The emission at other frequencies is opaque, resulting from the spectral index's positive values and the polarization properties discussed in \autoref{sec:polar}. 

The source was observed with the VLBA also in 2007 at three separate epochs. In a future study, we will be able to estimate the Doppler factor from the kinematics of the jet using multi-epoch VLBA observations and compare it with the estimates obtained from the variability in \autoref{sec:doppler}. One option can be that the component `J' size and its observed variability time scale are consistent for certain Doppler factor values. Finding the compact core `C' as the main source of variability might be another option.

\subsection{Modelling of the quasar structure and origin of the brightest jet feature}
\label{s:modeling}

\begin{figure}
	\includegraphics[width=\linewidth,trim={1.5cm 1.8cm 2cm 2cm}]{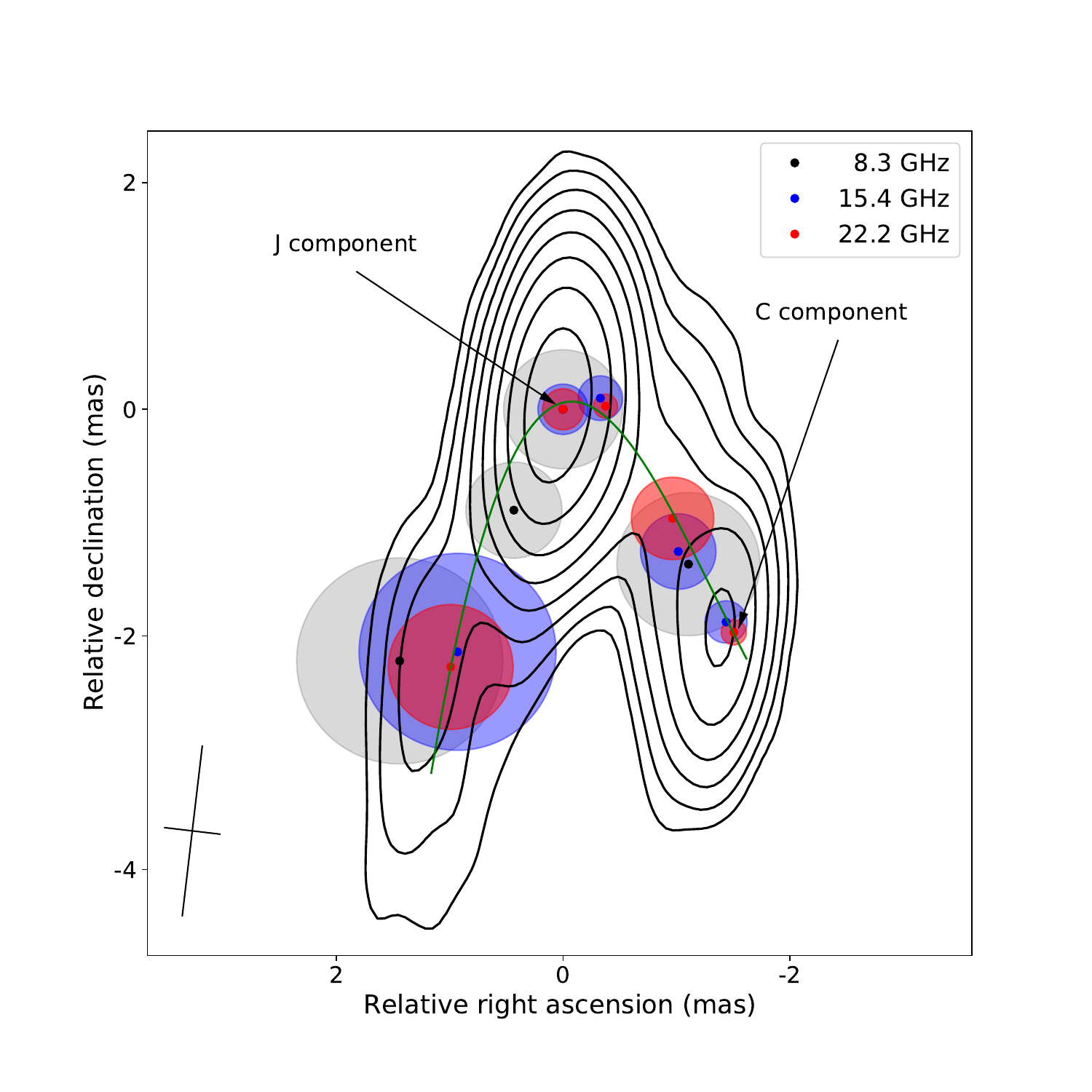}
    \caption{Ridge line drawn through the locations of the peaks of Gaussian components taken at 22.2~GHz. The full intensity contours are shown for the Stokes $I$ map at 22.2~GHz and start at the 2$\sigma_I$ level with the $\times2$ step. The peak positions and the sizes of the Gaussian components obtained at 8~GHz, 15~GHz and 22~GHz are shown in dots and circles. The arrows show the dominant jet component and the component associated with the core. The beam is shown in the bottom left corner at the FWHM level.}
    \label{fig:ridge_line}
\end{figure}

\begin{figure*}
	\includegraphics[width=\linewidth,trim={4cm 1cm 1cm 2cm},clip]{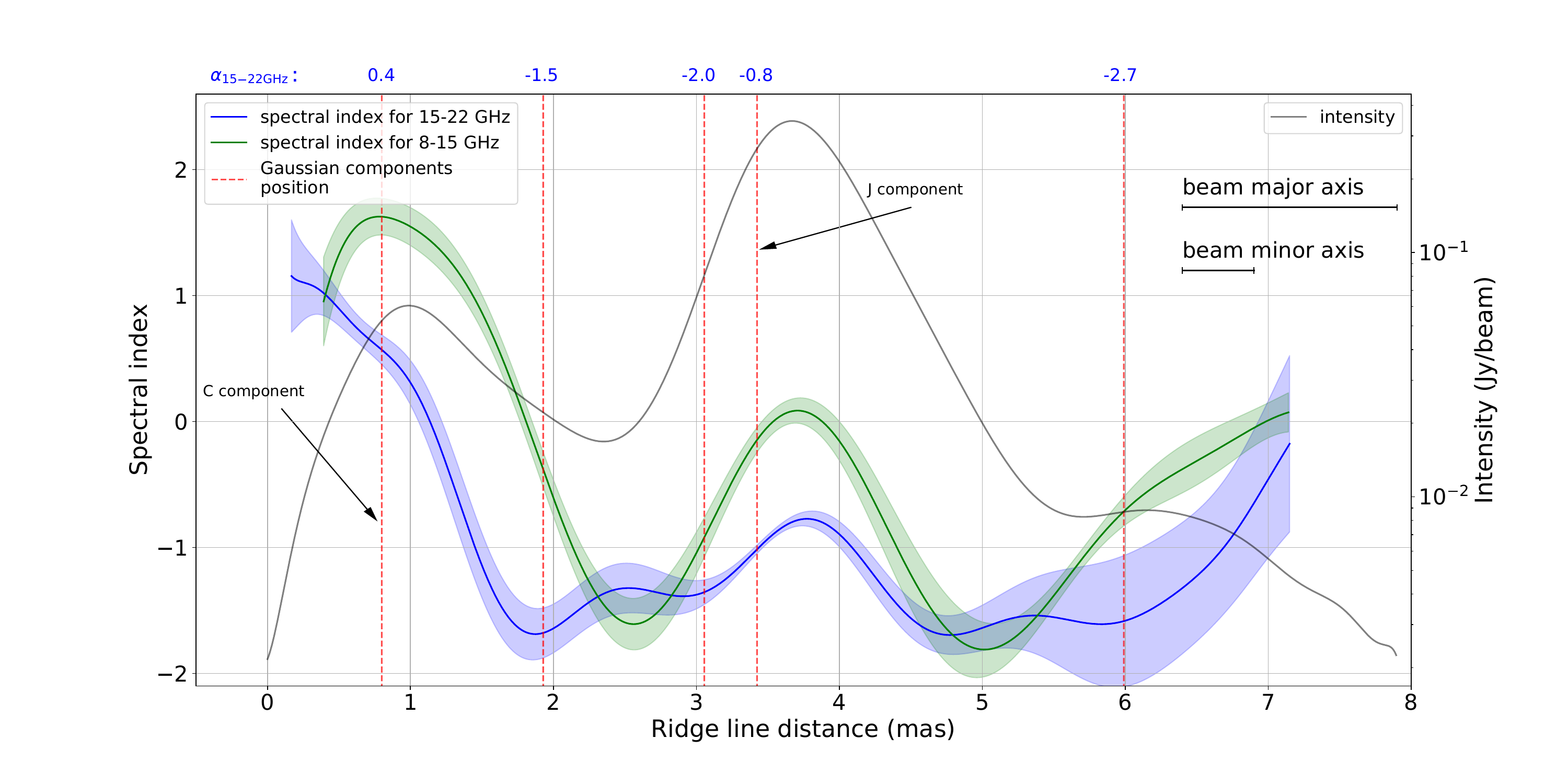}
    \caption{Distribution of the spectral index along the ridge line (\autoref{fig:ridge_line}). The spectral index dependencies for 8$-$15~GHz (green), 15$-$22~GHz (blue) and the distribution of the 22~GHz total intensity (gray). The positions of the Gaussian components are marked in red. The arrows show the dominant jet component and the component associated with the core. At the top right corner, there are major and minor axes of the beam from 22.2~GHz. The error of the spectral index distribution was taken from the spectral index error maps (\autoref{fig:spi_maps}). Spectral indices calculated from cross-identified components at 15 and 22~GHz are shown at the top of the plot in blue above the corresponding positions of these components. They were calculated using the corresponding total flux densities of these components (see \autoref{tab:gauss_comp}). }
    \label{fig:spi_ridge_line}
\end{figure*}

The Stokes $I$ maps reveal the core-jet structure of the quasar. To model that structure and study the spectral index distribution along the jet flow direction, a set of circular Gaussian components at each frequency separately was fitted to the calibrated visibility data in difmap. In \autoref{fig:ridge_line}, we plot the smoothed third-degree polynomial curve, ridge line, connecting the peaks of the Gaussian components obtained at 22.2~GHz. This curve characterises the jet trajectory projected on the sky. In \autoref{sec:spi_maps}, we performed a spectral index analysis to search for and identify the opaque core region, which turned out to be resolved from the jet at the highest observing frequencies. We determined the direction of the jet flow. Furthermore, we show in \autoref{tab:gauss_comp} the parameters of the components at the three highest frequencies where the structure of the source is visible before aligning on the optically thin component.

\begin{table}
	\centering
    \caption{Circular Gaussian model fitting results. 
    \label{tab:gauss_comp}}	
    \begin{threeparttable}
	    \begin{tabular}{ccccrl} 
		    \hline
            Frequency & Model & Flux density & Distance  &  P.A. &  FWHM \\
            (GHz) & $\chi^2_\mathrm{red}$ &(Jy) & (mas) & (deg.) & (mas)\\
            (1) & (2) &(3) &(4) &(5) & (6) \\
            \hline
            22.2 & 1.7 & 0.07  &  2.46  &  $-$143 & 0.22\tnote{C} \\   
             & & 0.44  &  0.05  &  139    & 0.35\tnote{J}\\   
             & & 0.07  &  0.34  &  $-$91  & 0.21 \\
             & & 0.03  &  2.49  &  156    & 1.09 \\  
             & & 0.04  &  1.35  &  $-$137 & 0.72 \\
             \hline
             15.4 & 2.4 & 0.06  &  2.35  &  $-$143 & 0.37\tnote{C}\\ 
              & & 0.59  &  0.06  &  90     & 0.44\tnote{J}\\   
              & & 0.15  &  0.29  &  $-$78  & 0.39 \\
              & & 0.08  &  2.36  &  156    & 1.72 \\
              & & 0.07  &  1.60  &  $-$142 & 0.66 \\
            \hline
             8.3 & 6.2 &0.19  &  1.80  &  $-$143 & 1.25\tnote{C}\\ 
              & & 1.12  &  0.05  &  $-$17  & 1.04\tnote{J}\\   
              & & 0.09  &  1.04  &  155    & 0.84 \\
              & & 0.14  &  2.68  &  148    & 1.80 \\
            \hline
        \end{tabular}
        \begin{tablenotes}
            \item The table shows: (1) the frequency of observation of the quasar, (2) the reduced chi-square $\chi^2_\mathrm{red}$ of the model, (3) the flux density from the component, (4) the radial distance of the component centre from the centre of the map, (5) the position angle of the centre of the component, (6) the FWHM size of the circular Gaussian component $\theta_\mathrm{G}$.
            
            \item [C] Component associated with the core.
            \item [J] The dominant jet component.
            See also \autoref{fig:ridge_line} and \autoref{fig:spi_ridge_line}.
        \end{tablenotes}
	\end{threeparttable}
\end{table}

We analysed the size-frequency dependence for the quasar. The following factors affect the observed sizes of the core and the jet features: (i) intrinsic properties which, in general, depend on synchrotron opacity, (ii) our ability to resolve a given studied feature from nearby jet structures, (iii) external scattering in our Galaxy. The galactic longitude and latitude of the quasar are $l=253.3^\circ$, $b=+11.8^\circ$, respectively. 
The measured core size at 15~GHz and 22~GHz depends on frequency as $\propto \nu^{-1}$ within the errors, following the conical synchrotron jet model \citep[][]{1979ApJ...232...34B}. This means that there is no significant scattering contribution, at least, at these frequencies.
At the same time, the dominant jet feature is very large~--- several times larger than typical AGN core sizes measured in this direction 
(Koryukova et al., in prep.). It shows $\propto \nu^{-1.8}$ dependence below 15~GHz, which might result from a combination of all three effects mentioned above. Note that strong scattering is possible but unlikely probable for these sky coordinates \citep{2015MNRAS.452.4274P}. 
It is difficult to separate the effects, and in this paper, we worked under an assumption that scattering does not affect the results significantly. Note that in our physical estimates, we mostly use the angular size measurements coming from the high frequencies.

We obtained the distribution of the spectral indices along the ridge line. The 22.2~GHz, 15.4~GHz and 8.3~GHz bands were used. At these frequencies, the jet structure can be properly traced due to the high enough resolution and transparency to radiation. The dependence of the spectral index on the distance along the ridge line is shown in \autoref{fig:spi_ridge_line}. The images were aligned, and the errors were calculated following \autoref{sec:spi_maps}.

We note that the values of the spectral index are the highest in the region of the core. They decrease with increasing the distance along the jet and then increase again at the dominant jet feature `J'. The spectral index $\alpha$ characterises the electron energy distribution. In the case of the simple homogeneous synchrotron source model $\alpha = \frac{(1-p)}{2}$, where $p$ indicates the energy distribution of electrons $N(E) =N_0E^{-p}$, $E$ is the electron energy. Therefore, we suggest that electrons are  accelerated in this region.

There are several viable explanations for the origin of the bright dominant region `J' in the jet. One of them is an increase of Doppler boosting of radiation at the bend of the jet \citep[e.g.][]{2003ASPC..299..117K}. Another one is the interaction of the jet with its environment \citep[e.g.][]{2000A&A...361..529A}. In the case of such an interaction, shock waves can arise, and the accelerated electrons can serve as a source of strong synchrotron radiation. Some sources were found with a similar bright jet component. For instance, the quasar 0923+392 (4C\,+39.25), which has a GPS-like spectrum and a similar spectral index distribution at parsec-scales, has been studied in several papers \citep[e.g.][]{1989A&A...211L..23M,1993ApJ...402..160A,1995AJ....110.2586G,2000A&A...361..529A}. The origin of the bright component in 0923+392 was interpreted as a shock wave propagating down the jet which curves its trajectory toward the observer. It is necessary to study the properties of the ambient medium surrounding the jet to determine which of the hypotheses is correct in our case. Since the incoming radiation is partially polarized, discerning between hypotheses can be done using the polarization measurements.

In case of an interaction with an ambient cloud, this cloud can be compressed and ionized, producing free-free absorption \citep[e.g.][]{1966AuJPh..19..195K}. This option should lead to a steeper rising spectrum below the peak frequency than in the case of synchrotron self-absorption. Some sources exhibit such behaviour \citep[e.g.][]{2000PASJ...52..209K}. Our observations demonstrate no direct indication of the presence of this effect. However, we cannot rule out this possibility as well. High-resolution ground-space observations at low frequencies could help us determine if free-free absorption is present in this source.

\section{Estimating magnetic field strength in the region of the dominant jet feature}
\label{sec:magnetic}

We used the formula derived by \citet{1983ApJ...264..296M} to estimate the magnetic field in the bright component:
\begin{equation}
    B = 10^{-5} b(\alpha) \theta^{4} \nu_{\mathrm{m}}^{5} S_{\mathrm{m}}^{-2} \left (  \frac{\delta }{1+z}\right )\,.
	\label{eq:magn_field_eq}
\end{equation}
In this expression, $B$ is the magnetic field in G, $\delta$ is the relativistic Doppler factor, $b(\alpha)$ is the coefficient which depends on the spectral index of optically thin radiation in the considered region \citep[see analytical derivation in][]{2019MNRAS.482.2336P}, $z$ is the redshift, $S_{\mathrm{m}}$ is the value of the flux density in Jy at the turnover frequency $\nu_{\mathrm{m}}$ in GHz. Note that $\theta = 1.8 \theta_G $ in mas is a reasonable estimate of the angular size of the considered spherical region, where $\theta_G$ is the size of the fitted Gaussian component \citep{1983ApJ...264..296M}. Initially, the expression was proposed by \citet{1963Natur.199..682S} to estimate the angular size of extragalactic radio sources in the case when the magnetic field values are known. It is assumed that the source is uniform, spherical and compact. 

With the aim to study the dominant feature `J', we modelled its spectrum by assuming it to be homogeneous, which allowed us to describe it, for both optically thin and thick regions, following e.g.\ \citet{1966Natur.211.1131V,1970ranp.book.....P}:
\begin{equation}
    I(\nu) = I_1\left ( \frac{\nu}{\nu_1} \right )^{2.5}\frac{1-\mathrm{exp}(-(\nu/\nu_1)^{\alpha_{\mathrm{thin}}-2.5})}{1-\mathrm{exp}(-1)}\,,
	\label{eq:homog_synchr_rad_eq}
\end{equation}
where $I_1$ and $\nu_1$ are the intensity and frequency corresponding to the optical depth $\tau_\mathrm{depth} = 1$, and $\alpha$ is the optically thin spectral index. We also allowed the spectral index values in the optically thick regime to be different from $+2.5$. In this way, we consider the possibility of non-uniformity of the source or a free-free absorption phenomenon in the ionized gas surrounding it \citep[e.g.][]{1966AuJPh..19..195K}. Therefore, we denote the optically thick index as $\alpha_{\mathrm{thick}}$ and, for the sake of clarity, we will use the notation $\alpha_{\mathrm{thin}}$ instead of just $\alpha$. It is convenient to rewrite \autoref{eq:homog_synchr_rad_eq} in terms of $I_{\mathrm{m}}$ and $\nu_{\mathrm{m}}$, where $I_{\mathrm{m}}$ is the maximum intensity in the spectrum, and $\nu_{\mathrm{m}}$ is the frequency where this maximum is reached \citep[e.g.][]{1970ranp.book.....P}:
\begin{equation}
    I(\nu) = I_{\mathrm{m}}\left ( \frac{\nu}{\nu_{\mathrm{m}}} \right )^{\alpha_{\mathrm{thick}}}\frac{1-\mathrm{exp}(-\tau_{\mathrm{m}}(\nu/\nu_{\mathrm{m}})^{\alpha_{\mathrm{thin}}-\alpha_{\mathrm{thick}}})}{1-\mathrm{exp}(-\tau_{\mathrm{m}})}\,.
	\label{eq:homog_synchr_rad_eq1}
\end{equation}
Here $\tau_{\mathrm{m}}$ is the optical depth at the frequency $\nu_{\mathrm{m}}$. 

We modelled the quasar structure in difmap with three Gaussian components, one of which being responsible for the bright jet feature, to find the frequency dependence of the flux density in the dominant jet feature. At first, the Gaussian components were taken from the fit at 15.4~GHz. Fixing their sizes, shapes and positions, we used them at all other frequencies following \citet[][]{2008evn..confE...9S}. That set was sufficient to adequately model the parsec-scale structure at 22.2~GHz, 15.4~GHz, 8.3~GHz and 4.8~GHz since the resolution at these frequencies was high enough. We established a bad fit criterion for the two remaining frequencies, 2.3~GHz and 1.5~GHz, due to the high value of $\chi^{2}>10$ of the model. Thus, we did not use the modelling results from these two lowest frequencies.

\begin{figure}
    \includegraphics[width=\linewidth,trim={1cm 0cm 2cm 1cm},clip]{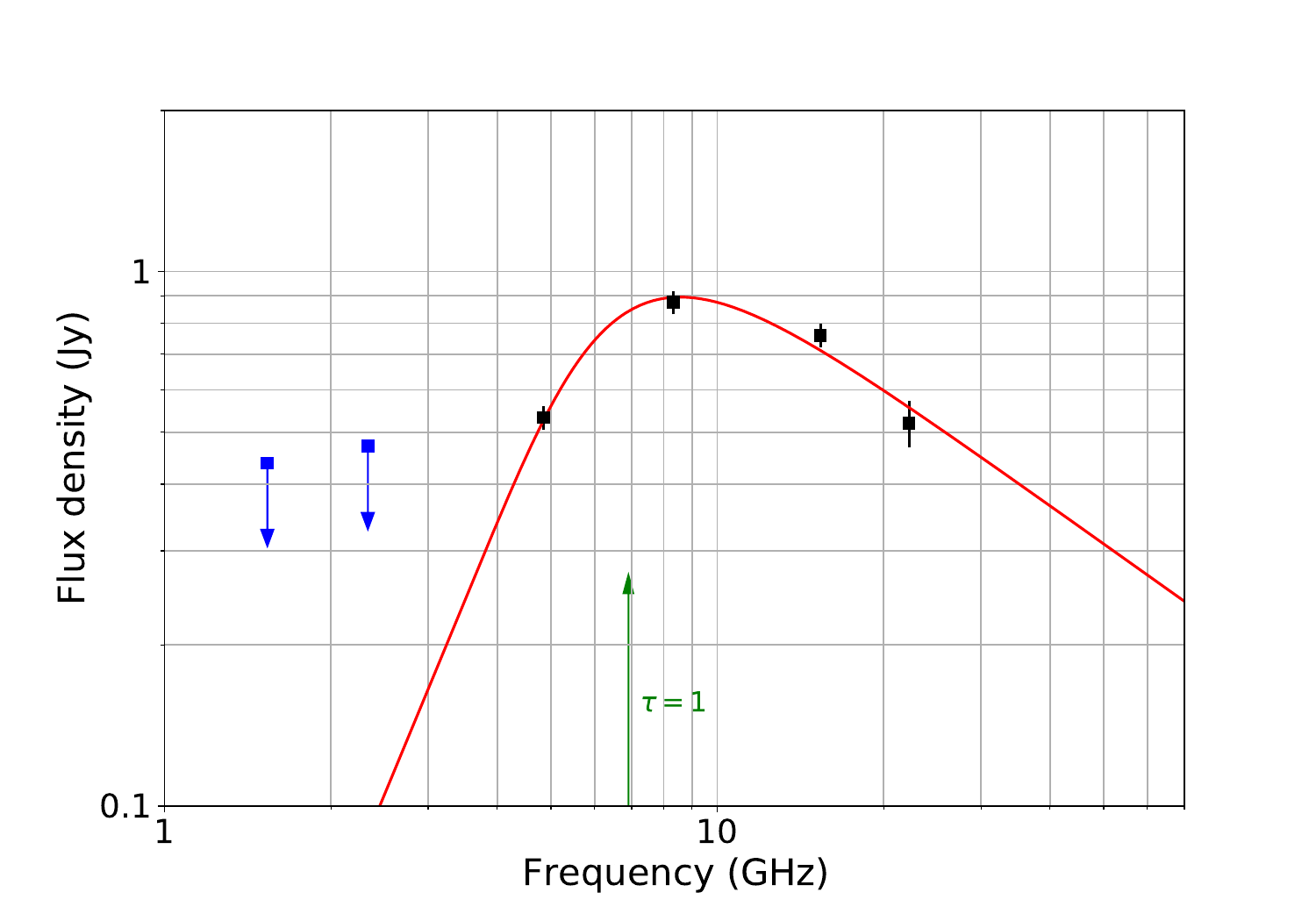}
	\includegraphics[width=\linewidth,trim={1cm 0cm 2cm 1cm},clip]{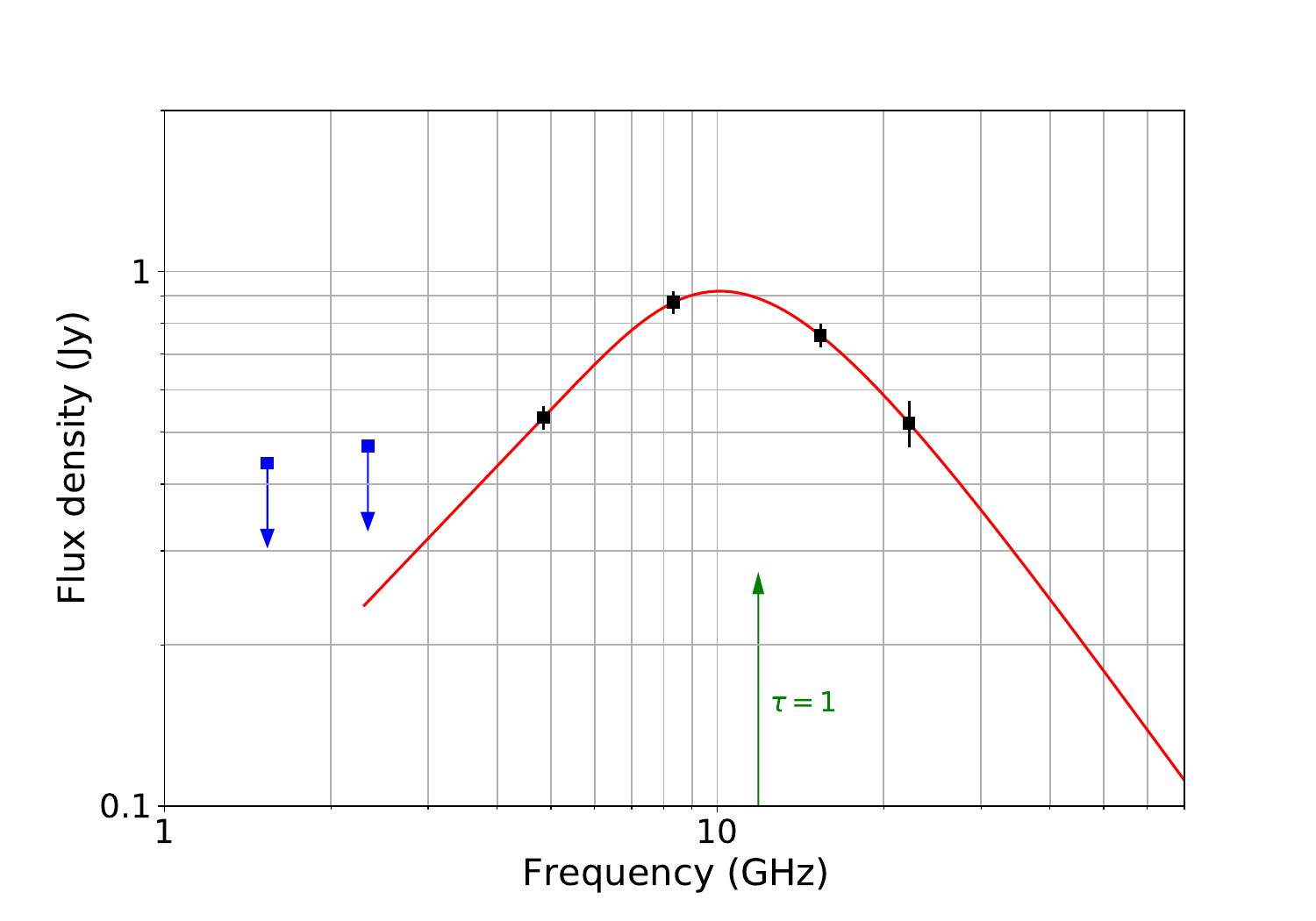}
    \caption{Frequency dependence for the flux density of the Gaussian component associated with the dominant jet feature. The points not used in the fitting to a synchrotron spectrum are highlighted in blue. The blue arrows show that these points correspond to the upper limits on the flux density. The green arrow shows the frequency corresponding to the optical depth $\tau_\mathrm{depth} = 1$. The top plot presents the model with the optically thick spectral index $\alpha_\mathrm{thick}$ = 2.5 in the case of a homogeneous synchrotron source, the bottom plot presents the model with an arbitrary one for the generalised approach (see \autoref{sec:magnetic}).}
    \label{fig:magn_field_est_fig}
\end{figure}

As anticipated above, we used both approaches to fit the flux density -- frequency data: for a homogeneous radiation source and generalised with an arbitrary value of the spectral index for the optically thick mode (see \autoref{fig:magn_field_est_fig}). The upper limits on the component flux density values at the two lowest frequencies are also shown in the figure, although they were not used for the modelling. We present the fitted parameters of the obtained models in \autoref{tab:magn_field_est_table}. 

\begin{table}
	\centering
	\caption{Fitting results for the spectrum of the dominant jet component `J' for the two considered models.}
	\label{tab:magn_field_est_table}
	\begin{tabular}{c|cccc} 
		\hline
        Model type & $\nu_{\mathrm{m}}$  & $\nu_1$  & $\alpha_{\mathrm{thick}}$ &  $\alpha_{\mathrm{thin}}$\\
        &(GHz) & (GHz) & & \\
        \hline
        Homogeneous & $8.5\pm0.4$ &  $6.9\pm0.5$  & 5/2  & $-0.73\pm0.16$ \\
        Generalized & 9.6  &  11.9  &  1.08 & $-1.42$ \\
    \hline
	\end{tabular}
	\begin{tablenotes}
            \item The table shows the values of the frequency $\nu_{\mathrm{m}}$, where the flux density maximum is reached, the frequency $\nu_1$,  where the optical depth becomes equal to 1, the spectral indices of thick and thin radiation.
        \end{tablenotes}
\end{table}

The errors of $\nu_{\mathrm{m}}$ and the optically thin spectral index $\alpha_{\mathrm{thin}}$ were derived using the variance–covariance matrix of the fit. Human-made decision on the number of components representing the overall structure has a significant effect on the VLBI-component size. This determines the model estimate of the size uncertainty. Thus, we modelled the quasar spectrum using a different number of components and found the dispersion of the brightest component size, with the aim to estimate the error of the circular Gaussian component size $\theta_G$. Eventually, the dominant component size was found to be $\theta_G = 0.54 \pm 0.03$ mas. 

As a result of this analysis, we obtained the estimate $B = (0.55 \pm 0.37)\delta$~G for the case of a homogeneous synchrotron source, and $B = 1.1 \delta$ for the generalised case.
For $\delta = 2.1 \pm 0.1$, taken from the variability Doppler factor estimates in \autoref{sec:doppler}, the value of the magnetic field in the dominant feature of the jet is of the order of 1~G.

Such a strong magnetic field indicates that the synchrotron losses at the brightest detail of the source are significant. The time during which an electron loses half of its energy for synchrotron radiation is $t_{1/2} \approx 10^{9}/(B^2\gamma_0)$ in seconds \citep[e.g.][]{2011hea..book.....L} , where $\gamma_0$ is the electron gamma factor, and $B$ is given in G. As an estimation taking $\gamma_0 = 10^{3}$, we find that the electron will lose most of its energy in about ten days. Since the bright detail is responsible for a strong change in the flux with time, and the quasar was observed for a few years before and after 2005, we can conclude that in this place of the jet, there should be a continuous re-acceleration of emitting relativistic particles. The strong magnetic field is an indication that due to the interaction of the jet with the surrounding environment, a recollimation shock wave is formed, and a bright emission appears from this region \citep[e.g.][]{2015AN....336..447F}.

\section{Polarization properties}
\label{sec:polar}

The complex linear polarization is defined as:
\begin{equation}
    \Pi = Q + iU = Pe^{2i\chi}\,,
	\label{eq:polar_eq}
\end{equation}
where $I$, $Q$ and $U$ are the Stokes parameters, $P$ is the linear polarization modulus $\Pi$ and $\chi$ is the electric vector position angle (EVPA). The noise level on the maps of the linear polarization was calculated as the averaged sum of $\sigma_U$ and $\sigma_Q$ noise levels of the CLEAN maps for the Stokes parameters $U$ and $Q$, respectively \citep[e.g.][]{2012AJ....144..105H}.

\begin{figure*}
	\begin{center}
    \resizebox{0.9\textwidth}{!}{%
    \includegraphics[width=0.45\linewidth, trim=1cm 1.cm 0.5cm 2cm,clip]{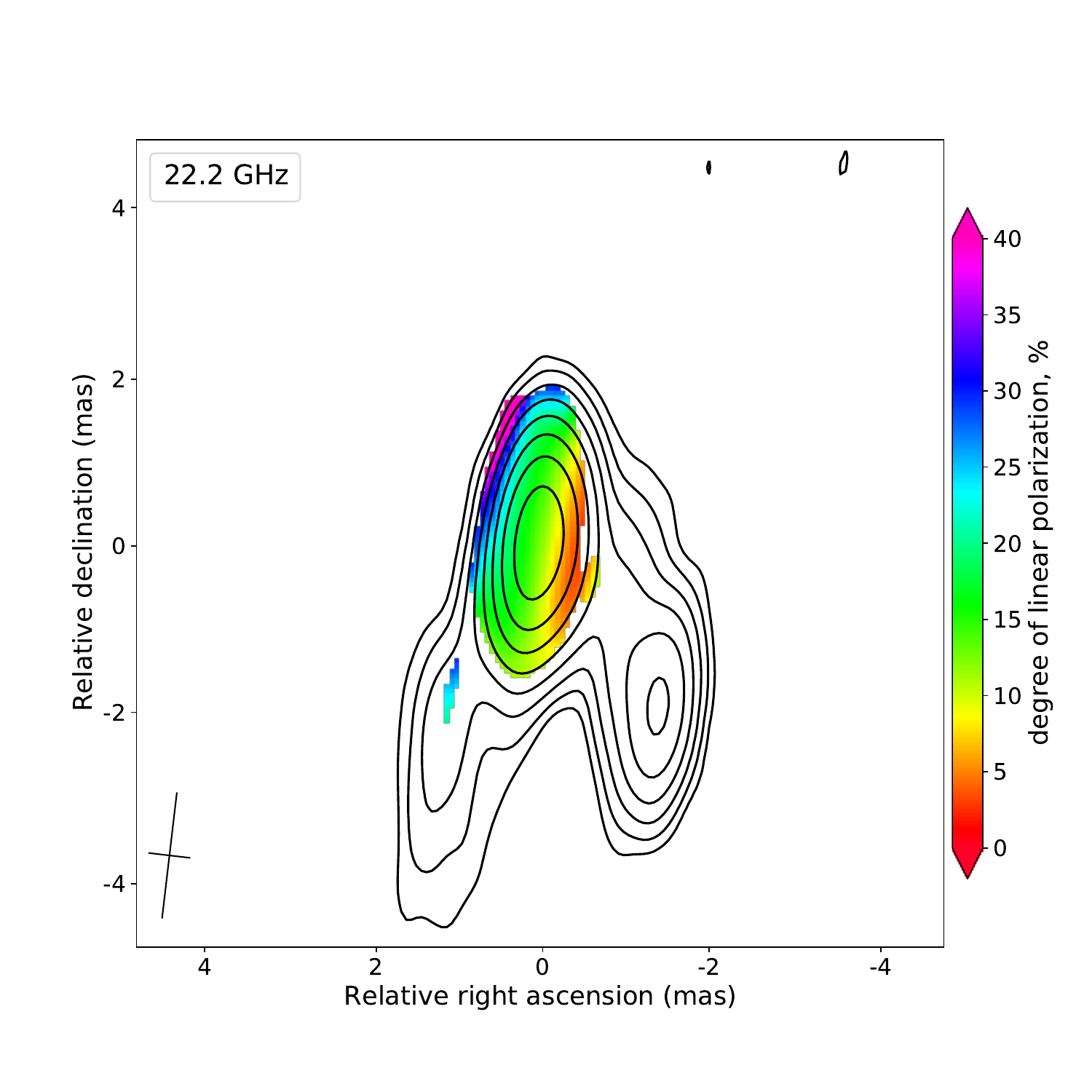}%
    \includegraphics[width=0.45\linewidth,trim=1cm 1.cm 0.5cm 2cm,clip]{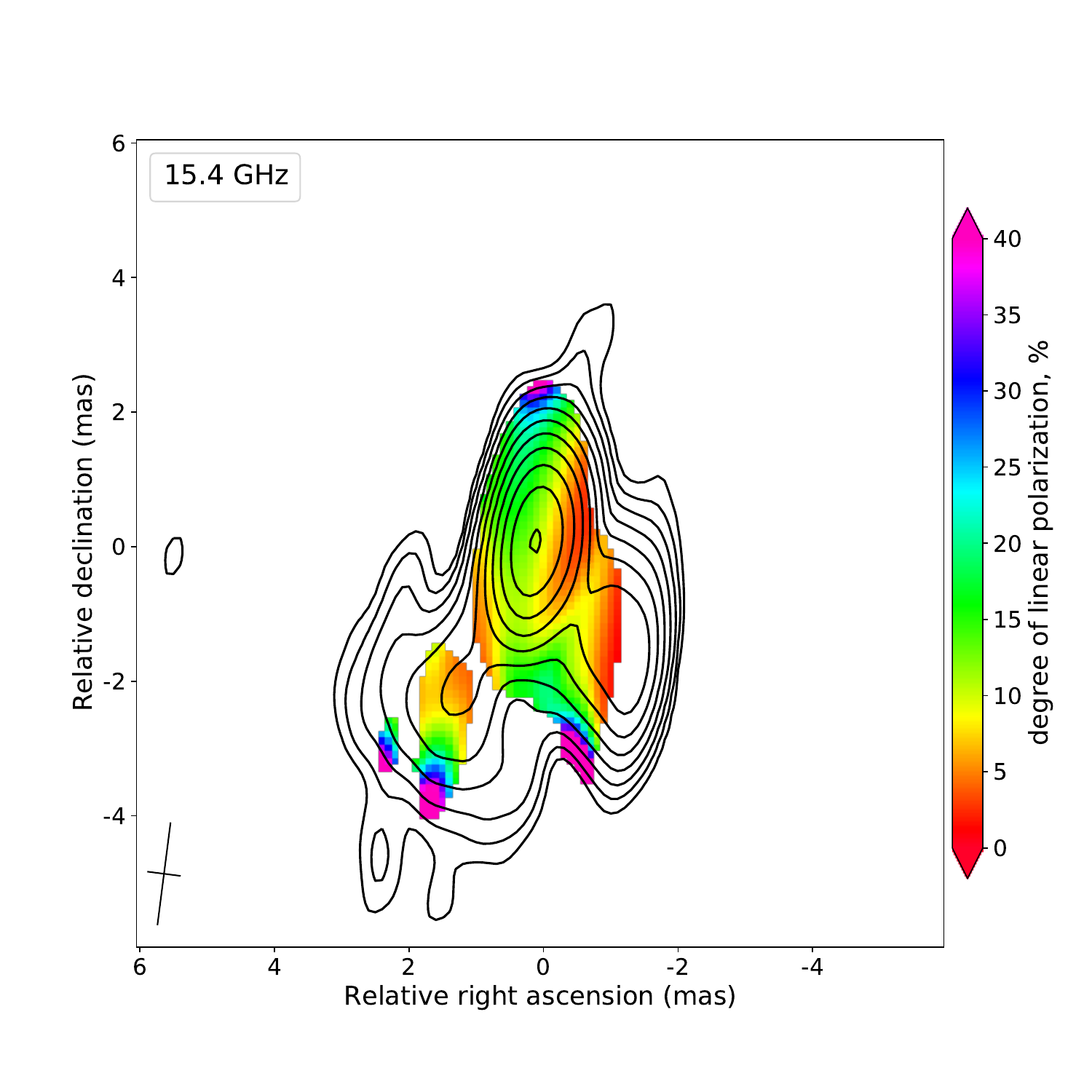}%
    }
    \resizebox{0.9\textwidth}{!}{%
    \includegraphics[width=0.45\linewidth, trim=1cm 1.cm 0.5cm 2cm,clip]{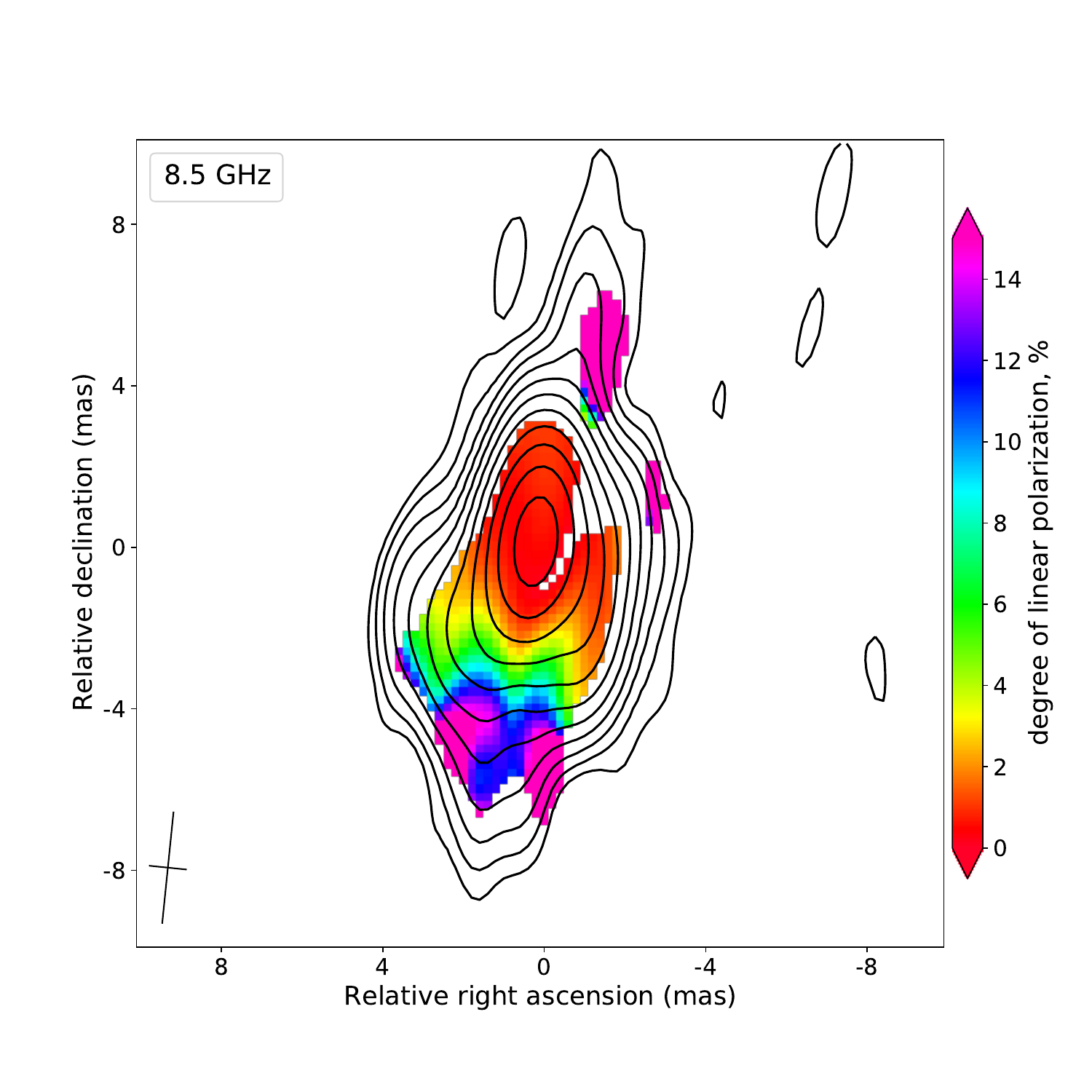}%
    \includegraphics[width=0.45\linewidth,trim=1cm 1.cm 0.5cm 2cm,clip]{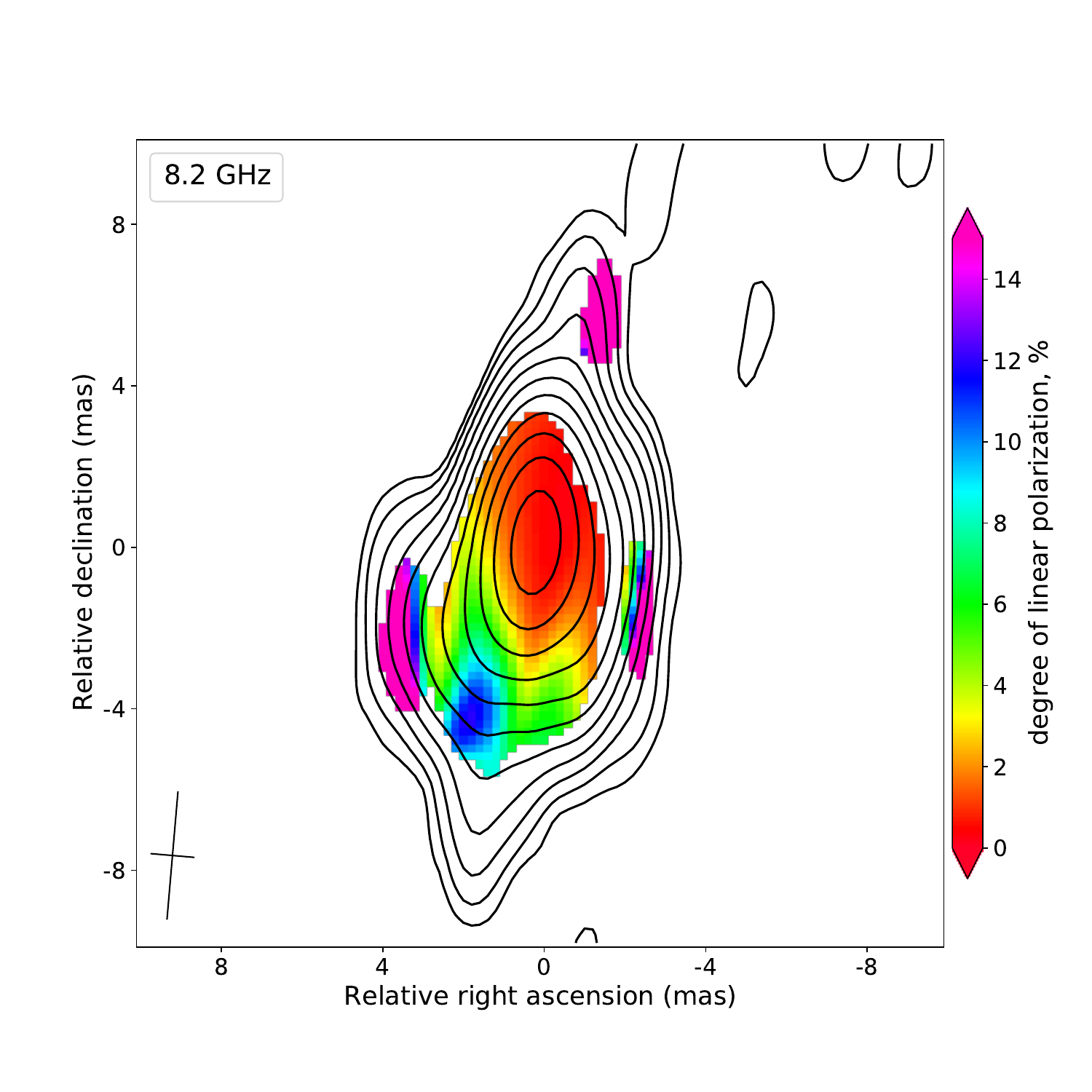}%
    }
    \end{center}
    \caption{Degree of linear polarization maps for 0858$-$279 at 22.2~GHz (top left), 15.4~GHz (top right), 8.5~GHz (bottom left) and 8.2~GHz (bottom left). The maps show the full intensity contours starting at the 2$\sigma_I$ level. The degree of linear polarization is shown within the linear polarization contours starting at the 3$\sigma_P$ level. It can be seen that the polarization degree decreases with increasing the wavelength due to depolarization effects. The restoring beam is shown as a cross at the FWHM level in the left bottom corner.}
    \label{fig:deg_lin_pol_fig}
\end{figure*}

One of the most useful ways to characterise the properties of radiation is the degree of its linear polarization, which is introduced as
\begin{equation}
    m = \frac{\sqrt{U^{2}+Q^{2}}}{I}\,.
	\label{eq:deg_lin_polar_eq}
\end{equation}
Due to the depolarization phenomenon \citep[e.g.][]{2012AJ....144..105H,2012MNRAS.421.3300O, 2017MNRAS.467...83K}, the polarization properties, including the degree of linear polarization, become wavelength dependent. Apart from that, they depend on whether the regime of radiation is optically thick or thin. Maps of the distribution of the degree of linear polarization were reconstructed at six frequencies to study depolarization. Here, we show the maps at the three highest frequencies (\autoref{fig:deg_lin_pol_fig}). Apparently, the degree of linear polarization decreases with increasing the wavelength. It can originate from depolarization caused by internal or external Faraday rotation or because the images are reconstructed with different beam sizes. Another reason for decreasing the degree is that the radiation regime changes from optically thin to optically thick with decreasing the frequency. ~\autoref{tab:lin_degree_table} shows the degree obtained over the region of the map which starts at the 3$\sigma_P$ level for the four highest frequencies. The degree at other frequencies turned out to be less than 1.5$^{\circ}$.

\begin{table}
	\centering
	\caption{Degree of linear polarization, $m$, obtained over the region of the map which starts at the 3$\sigma_P$ level for four highest frequencies.}
	\label{tab:lin_degree_table}
	\begin{tabular}{lc} 
		\hline
		Frequency & $m$ \\ 
		(GHz) & ($\%$)  \\
		\hline
		22.2 & 10   \\   
        15.4 & 8.1  \\
        8.5 & 0.5   \\
        8.2 & 1.1   \\
		\hline
	\end{tabular}
\end{table}

In the case of external Faraday rotation, when the rotation of the plane of polarization occurs in the medium surrounding the jet, the EVPA angle linearly depends on the squared wavelength:
\begin{equation}
    \chi = \chi_0 + \mbox{RM} \lambda^{2}\,,
	\label{eq:evpa_eq}
\end{equation}
where RM is the rotation measure \citep{1966MNRAS.133...67B} in rad/m$^{2}$. 
The rotation occurs in three main regions: near the jet in the distant active galaxy, in the intergalactic and galactic media. The intergalactic rotation \citep[e.g.][]{2010ApJ...723..476A} was not considered in this work since it is poorly studied. Apart from that, since the intergalactic concentration of particles is low, it should not affect the RM values significantly. The rotation in our Galaxy was investigated, and the RM values were obtained in various works \citep[e.g.][]{2009ApJ...702.1230T,2011ApJ...728...97V,2012ApJ...757...14J,2012A&A...542A..93O}. The value of the galactic rotation measure for the studied quasar is equal to $85\pm3$~rad/m$^{2}$. It was taken from the distribution map of the median value of RM as a function of galactic coordinates \citep{2009ApJ...702.1230T}. Later in the analysis, this value is subtracted to leave only the rotation measure in the medium surrounding the jet for our analysis.

For polarization imaging, the lower frequency bands were divided into two sub-bands of 8.2~GHz \& 8.5~GHz, 4.6~GHz \& 5.0~GHz, 2.3~GHz \& 2.4~GHz, and 1.4~GHz \& 1.7~GHz. The Faraday RM maps were reconstructed for the frequency ranges 1.4$-$2.4~GHz, 2.3$-$5.0~GHz, 8.2$-$15.4~GHz, and 15.4$-$22.2~GHz due to the changes in the optical depth properties and different image resolutions. 
The RM value was calculated as the slope of the linear dependence of EVPA on $\lambda^{2}$ for each pixel of the overlaid maps.
For every frequency range, all the corresponding maps were reconstructed with the same beam and pixel size taken from the lowest frequency of the range. The $\chi^{2}$ of the fitted EVPA -- $\lambda^2$ dependency was minimized to resolve the $n\pi$ ambiguity.

The process of overlaying is carried out in the same way as for the spectral index maps aligning on the Gaussian component `J'. We visually inspected resulting spectral index maps in the indicated frequency intervals and did not observe any artificial gradients. 
We used only those pixels where the modulus of linear polarization was greater than 3$\sigma_P$ and blanked the poor $\lambda^{2}$-fit regions based on a $\chi^{2}$ criterion. We utilised the corresponding 95$\%$ confidence limit from a $\chi^{2}$ distribution. 

\begin{figure}
    \centering
	\includegraphics[width=\linewidth,trim=1cm 2cm 0cm 2.5cm]{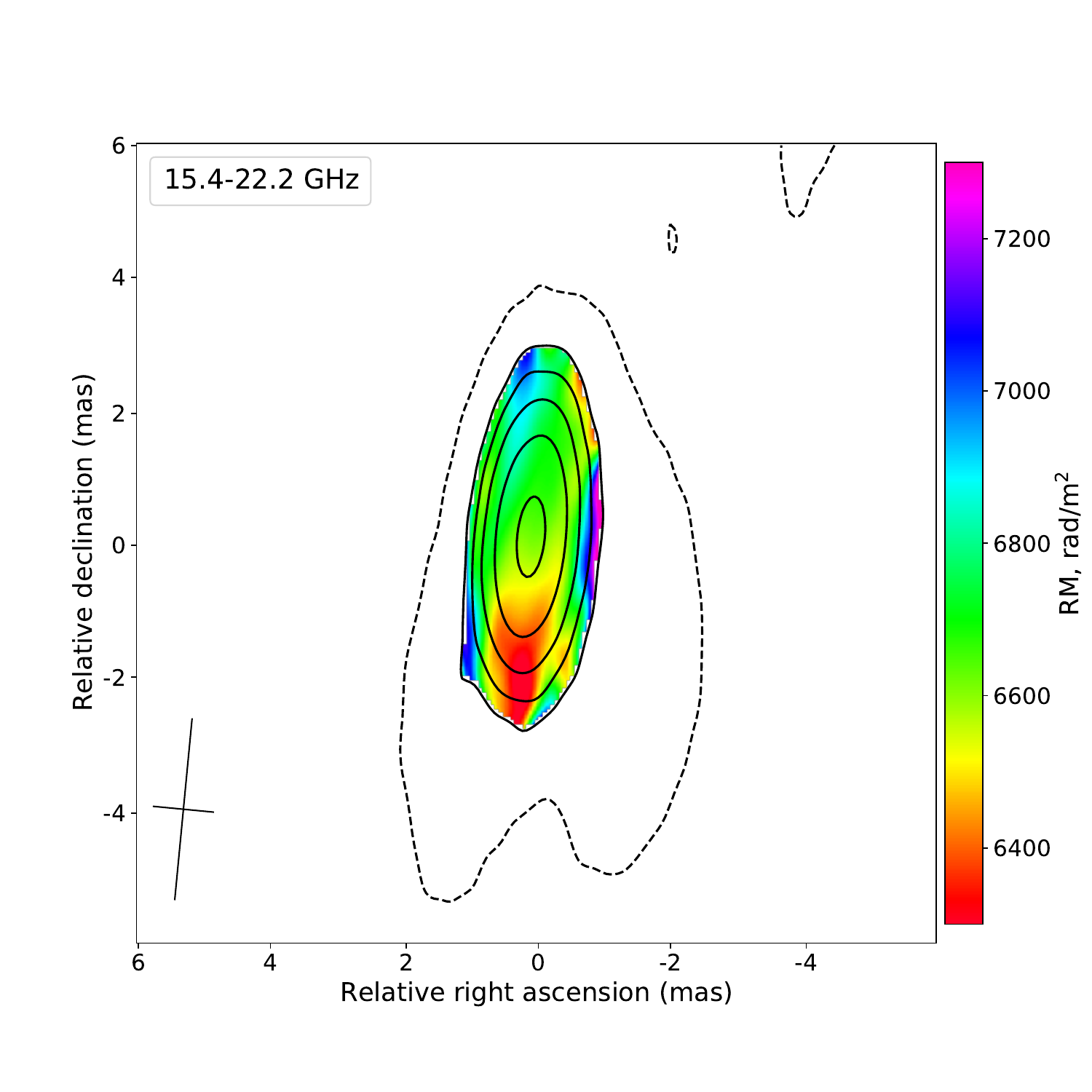}
    \caption{Faraday RM map for the frequency interval 15.4$-$22.2~GHz. The map shows the full intensity contour (dotted line) at 2$\sigma_I$ taken from the Stokes $I$ map for the highest frequency of the interval. The RM is shown within the linear polarization contours starting at the 3$\sigma_P$ level with the $\times2$ steps for the highest frequency as well. The restoring beam is shown as a cross in the bottom left corner at the FWHM level. The obtained values for this interval are in good agreement with values for lower frequency intervals (see \autoref{tab:polarization_table}).}
    \label{fig:RM_map_fig}
\end{figure}

\begin{figure}
    \centering
	\includegraphics[width=0.95\linewidth,trim=1cm 1.5cm 0cm 2cm]{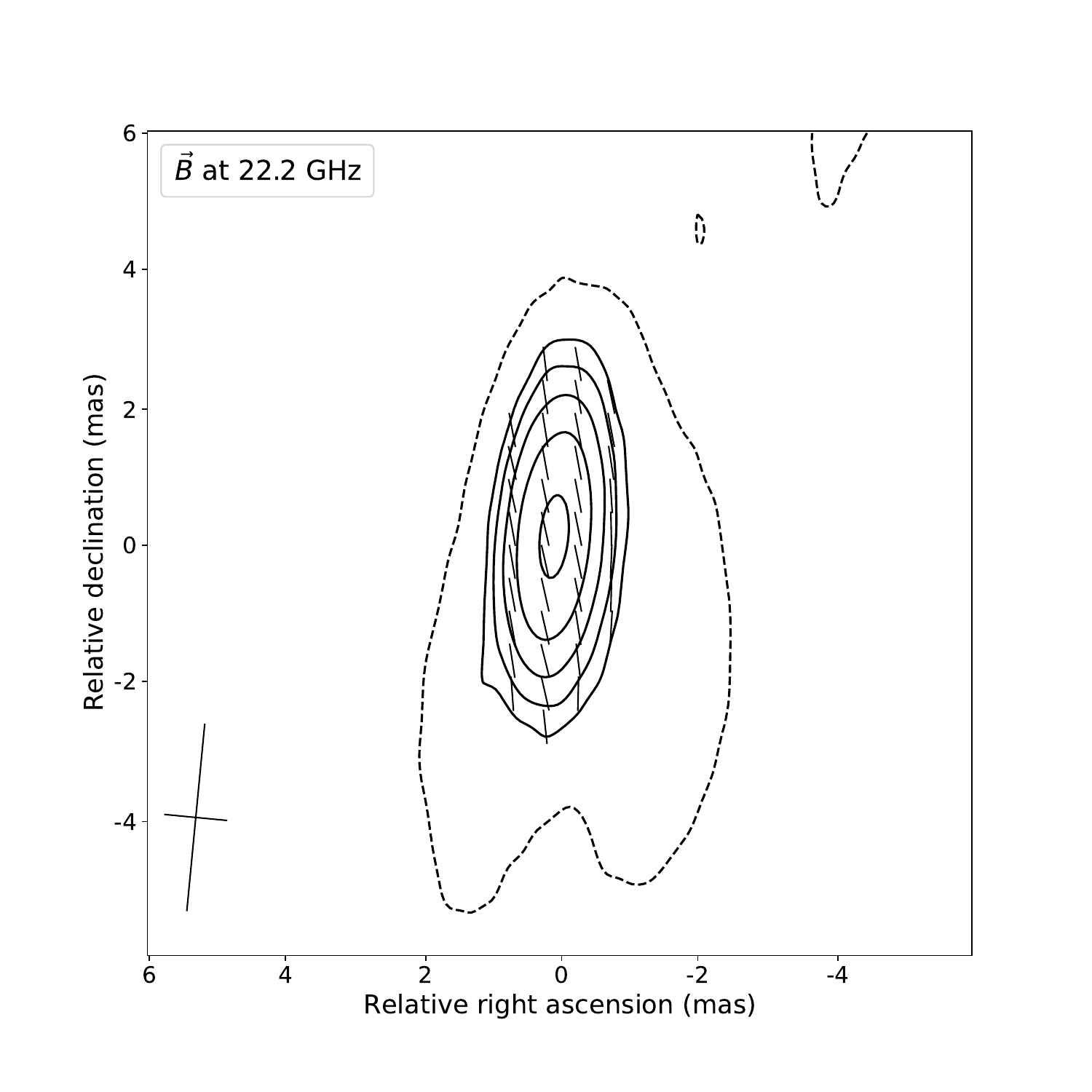}
    \caption{Structure of the magnetic field of the quasar at 22.2~GHz. See details in \autoref{tab:polarization_table}.
    The map shows the outer contour (dotted line), taken from the maps of the total intensity at the 2$\sigma_I$ level.
    The magnetic field direction is shown within the linear polarization contours starting at the 3$\sigma_P$ level with the $\times2$ steps. 
    The restoring beam is shown as a cross in the bottom left corner at the FWHM level.}
    \label{fig:magn_struc_fig}
\end{figure}

The EVPA error was calculated using the following formula:
\begin{equation}
    \sigma_{\mathrm{total}} = \sqrt{\sigma_\mathrm{EVPA}^{2}+\sigma_{\mathrm{Dterm}}^{2} +\sigma_\mathrm{VLA}^{2}}\,,
	\label{eq:evpa_err_eq}
\end{equation}
where $\sigma_\mathrm{EVPA}$ is the error related to the noises in the maps of the $U$ and $Q$ Stokes parameters, $\sigma_{\mathrm{Dterm}}$ is the error related to the polarization leakage parameters, with the error being taken equal to 0.2~percent of the total intensity \citep[e.g.][]{2012AJ....144..105H}, $\sigma_\mathrm{VLA}$ is the absolute EVPA calibration error obtained from the VLA data. 

Here, we present the Faraday rotation measure (RM) map for the interval 1.4$-$2.4~GHz (\autoref{fig:RM_map_fig}). Since there are only two frequencies for the highest range, the EVPA ambiguity problem was solved by minimizing the modulus of RM shifting EVPA at one frequency by $n\pi$. The not shown 1.4$-$2.4~GHz and 2.3$-$5.0~GHz RM maps look very similar with the polarized region showing effectively one RM value for the dominant component `J' since the pixels are codependent and it is unresolved at the lowest frequencies. 
We underline that RM values estimated from the frequency ranges shown in \autoref{tab:polarization_table} agree very well. This strongly confirms the two-frequency 15$-$22~GHz RM image and the determined $n\pi$ values.
The results we obtained for the 8$-$15~GHz interval show a clear non-$\lambda^{2}$ behaviour for most pixels. This is explained by changing opacity conditions at this frequency interval (\autoref{fig:magn_field_est_fig}) and consequently expected $90^\circ$ EVPA flip. Therefore, we do not utilise the 8$-$15~GHz interval for our analysis. 

\begin{table*}
	\centering
	\caption{Faraday rotation and direction of the magnetic field in the dominant jet feature `J'.}
	\label{tab:polarization_table}
	\begin{threeparttable}
	\begin{tabular}{rcccccc} 
		\hline
		Frequency & RM &  $B_\mathrm{maj}$ & $B_\mathrm{min}$ & $B_\mathrm{PA}$ & $\pi/2$  & Magnetic field  \\
		(GHz) & ($\times 10^3$ rad/m$^2$) & (mas) & (mas) & (deg) & correction & direction (deg) \\
		\hline
          1.4 & $6.30 \pm 0.11$ & 78 & 23 & 165 &  N & \dots  \\   
          1.7 &                 & & & &  N & \dots \\ 
          2.3 &                 & & & &  N & $170 \pm 99$  \\   
          2.4 &                 & & & &  N & $153 \pm 90$ \\  
        \hline
          2.3 & $6.31 \pm 0.12$ & 35 & 9.4 & 171 &  N & $153 \pm 102$  \\ 
          2.4 &                 & & & &  N & $166 \pm 123$  \\ 
          4.6 &                 & & & &  N & $160 \pm 30$  \\   
          5.0 &                 & & & &  N & $161 \pm 25$ \\
        \hline
          15.4 & $6.69 \pm 0.27$ & 2.7 & 0.9 & 175 &  Y & $191 \pm 3$  \\   
          22.2 &                 & & & &  Y & $191 \pm 3$ \\ 
        \hline
	\end{tabular}
	\begin{tablenotes}
            \item RM and magnetic field direction were averaged over 9 pixels at the location of the map containing the brightest pixel. The table is divided into three parts. The top part presents the results for RM obtained in the frequency range of 1.4$-$2.4~GHz, the middle part contains the results for 2.3$-$5.0~GHz and the bottom part present the results for 15.4$-$22.2~GHz. $B_\mathrm{maj}$ and $B_\mathrm{min}$ are the major and minor axes full width at half maximum (FWHM) of the restoring beam, respectively. $B_\mathrm{PA}$ is the position angle of the beam. The sixth column indicates whether a $90^\circ$ correction was introduced in the magnetic field in the case of optically thick synchrotron radiation (Y~--~yes, N~--~no). The magnetic field direction values for 1.4~GHz and 1.7~GHz are not shown because the error is greater than $180^\circ$.
            \end{tablenotes}
    \end{threeparttable}
\end{table*}

Using the obtained RM maps, it is feasible to reconstruct the intrinsic EVPA distribution structure in the jet. First, we correct the rotation. It is also necessary to take into account that in the optically thin mode of synchrotron radiation, the magnetic field is perpendicular to the EVPA. As shown in~\autoref{fig:spi_maps} and \autoref{fig:magn_field_est_fig}, the radiation occurs in the optically thin regime at 22.2~GHz and 15.4~GHz. Therefore, we corrected the EVPA by $90^\circ$ for these two highest frequencies to obtain the magnetic field direction. The magnetic field direction map at the highest frequency is presented in \autoref{fig:magn_struc_fig}. We show the image parameters in \autoref{tab:polarization_table}. We used Monte Carlo simulations to calculate the errors of the RM values. It is important to note that even small errors in RM result in strong uncertainty in the magnetic field direction at low frequencies. 
Apart from that, we estimated the EVPA rotation inside separate sub-bands since we obtained high RM values. The maximum value obtained at the lowest frequency turned out to be slightly smaller than $180^\circ$. Therefore, the angle averaging inside the 8~MHz-wide frequency channels did not affect the reconstructed rotation measure maps, although depolarization could have been significant.

As a result, there is a general trend in the direction of the magnetic field estimated at all frequencies. This confirms that the RM values and the $90^\circ$ EVPA corrections are robust.
We note that the direction of the magnetic field is perpendicular to the jet propagation direction after the bent and aligned with the compressed isophotes. The latter can be a consequence of the shock wave, as interpreted in, e.g. \citet{2002ApJ...568..639P}, in which hot-spots in Compact Symmetric Objects amplifying the magnetic field perpendicular to the shock front were identified. The former is known to be the case from plasma theory \citep[e.g.][]{2013arXiv1301.5572S}, and it has also been shown to be the expected situation in relativistic jets \citep[e.g.][]{1993A&A...274...55G,1994A&A...284...51G,1994A&A...292...33G,2016ApJ...831..163M,2018ApJ...860..121F,2021A&A...650A..60M,2021A&A...650A..61F}. Thus, both features point towards the presence of a shock wave in this region.

Such high RM values are not typical of the jet regions \citep[e.g.][]{2012AJ....144..105H,2017MNRAS.467...83K}, but some sources were reported to have extreme values \citep[e.g.][]{2002ApJ...566L...9Z}. If we roughly assume that the cloud magnetic field is 10$-$1000 $\mu$G, and its characteristic size is 1 pc, it is possible to estimate the electron density in the cloud $n_e \approx 10^2-10^4$ cm$^{-3}$. This estimation is consistent with narrow line region densities \citep[e.g.][]{2018A&A...618A...6K}, although they are mostly found in nearby galaxies. 





\section{Dominant jet feature origin and magnetic field}
\label{sec:origin}

\begin{figure}
    \centering
	\includegraphics[width=\linewidth,trim=0cm 8cm 0cm 0cm]{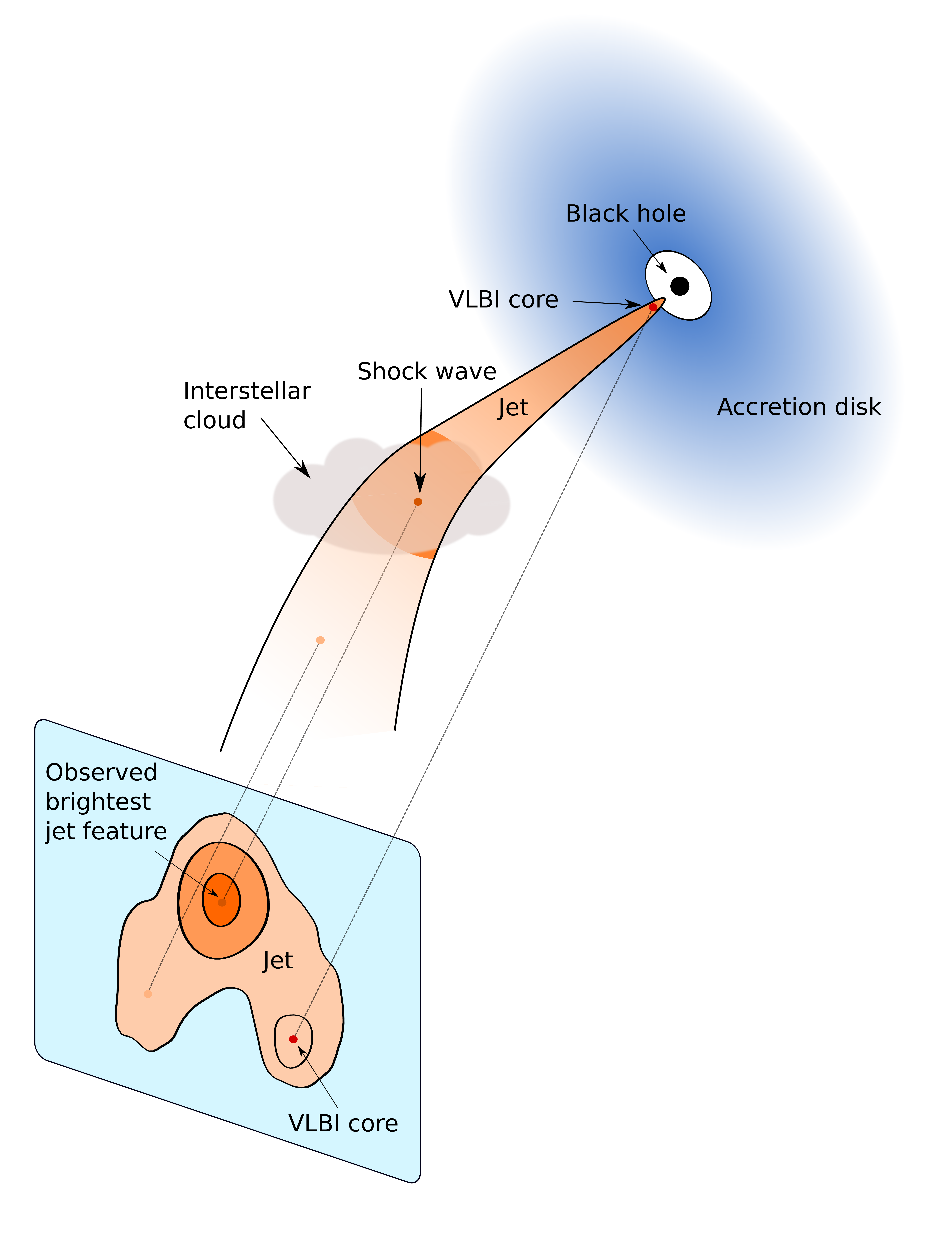}
	\caption{Schematic drawing which demonstrates the jet-cloud interaction hypothesis. 
	}
    \label{fig:0858_scheme}
\end{figure}

There are several options to explain the observed bright jet region. Since there is an apparent bend, the jet is likely to interact with a dense cloud of interstellar medium gas. Such interaction can lead to a jet propagation deviation and the occurrence of a shock wave. All the obtained results seem to support this idea. In the case of a shock wave, 1) the emission of electrons becomes brighter, 2) the direction of the magnetic field is perpendicular to the jet propagation direction after the observed bent and 3) the degree of linear polarization is relatively high as well, as a consequence of the magnetic field being highly ordered and amplified at the discontinuity. Aside from that, many works \citep[e.g.][]{2012AJ....144..105H,2017MNRAS.467...83K} have been devoted to study RMs of AGN. The RM values obtained in our study are unusually high for features being so far away from the core. They are yet another evidence that the jet pierces a dense cloud of interstellar matter and interacts with it. This cloud should be, at least partially, crossed by light rays in our line of sight towards the jet for a reasonable expectation of a small viewing angle \citep[e.g.][]{2004ApJ...609..539K,2009AJ....137.3718L}, which would favour rotation. 
We estimated the magnetic field strength in the dominant jet feature (\autoref{sec:magnetic}), assuming a homogeneous synchrotron source model with a randomly oriented magnetic field. Note that this is not consistent with the ordered magnetic field at the shock front scenario discussed here. We expect that the estimate is valid for an order of magnitude. 
Such high values are not typical of undisturbed jets. Therefore, this is another indication in favour of the proposed jet-cloud interaction hypothesis which we present in the schematic drawing (\autoref{fig:0858_scheme}).  

The second alternative is the radiation amplification by Doppler boosting since the jet viewing angle changes at the apparent bend point. However, it is difficult to justify this possibility unless the precise jet kinematics of the source is known.

Another option is a very young source with a hotspot. The southeast emission would then be caused by the fast shocked jet plasma, which forms the backflow. Although backflows can be fast \citep[e.g.][]{2019MNRAS.482.3718P}, they are unlikely to reach the velocity close to the speed of light to be beamed, and it would mainly propagate in the direction far from the viewing angle, so this option is also improbable as compared to a strong jet-ambient interaction.

If we assume that the standing shock wave produced by such an interaction is the origin of the bright region, we can estimate the pre-shock magnetic field. As long as the source is compact, the upstream speed should be close to the speed of light. Assuming that the degree between the shock and the flow is $90^\circ$, the upstream flow slows down to $c/3$ \citep[e.g.][]{2012rjag.book..245B}, where $c$ is the speed of light. Using the jump conditions $\Gamma_2 B_2 = \Gamma_1 r B_1$, where index~1 shows the upstream values, index~2 shows the downstream values, $r$ is the compression ratio, with $r = 3$ as discussed above, and $\Gamma_2$ is close to 1 due to deceleration at the shock. Thus, the magnetic field is amplified by $3 \Gamma_1$. 

Future studies of multi-epoch observations of this quasar will allow us to evaluate the kinematics parameters of a jet. For now, we use the rough approximation $\Gamma \sin \theta = 1$ and $\delta = 2.1 \pm 0.1$ found in \autoref{sec:doppler}, where $\theta$ is the angle the jet is viewed at. One can obtain that the parameter $\Gamma\approx 2$. Therefore, taking the value found in \autoref{sec:magnetic}, we can calculate the pre-shock magnetic field at the bright detail region, which turns out to be equal to $B_1 \approx 0.2$~G.

Applying the magnetic flux conservation, one can obtain the core region magnetic field value \citep[e.g.][]{2019A&A...631A..49K}. We expect the magnetic field to be toroidally dominated and neglect the poloidal component of the field. Therefore, the core magnetic field $B_\mathrm{core} = B_1 r_1/r_2$, where $r_1$ and $r_2$ are the sizes of the bright detail component and the core component, respectively. The core emission is opaque at 22.2~GHz and 15.4~GHz. That is why its size depends on the frequency as $\nu^{-1}$. Considering the highest frequency (\autoref{tab:gauss_comp}), we can find the lower limit for the core field $B_\mathrm{core} \approx 0.3$~G.

There is another approach to find the core magnetic field. It is based on the apparent core shift effect \citep{1979ApJ...232...34B,1981ApJ...243..700K,1998A&A...330...79L}. We used the Gaussian components from \autoref{tab:gauss_comp} and the same image alignment procedure based on the bright detail position to derive the frequency-dependent core shift. We modelled the quasar by a different number of components to estimate the measurement error and found the dispersion of the radial distance of the component centre. The value for the 22.2$-$15.4~GHz core shift turned out to be $\Delta r_\mathrm{core}= 0.06 \pm 0.03$~mas. The direction of the core shift vector coincides with the inner jet direction within the errors. Thus, we managed to calculate the core shift measure defined in \citet{1998A&A...330...79L} as:
\begin{equation}
    \Omega_{r\nu} = 4.85 \times 10^{-9} \frac{\Delta r_\mathrm{core} D_L}{(1+z)^2} \frac{\nu_2\nu_1}{\nu_2-\nu_1} \mathrm{pc\,GHz}\,,
	\label{eq:core_shift_measure_eq}
\end{equation}
where $\nu_1$ and $\nu_2$ are the considered frequencies in GHz, $D_L$ is the luminosity distance in parsecs and $\Delta r_\mathrm{core}$ is measured in mas. The final value $\Omega_{r\nu} = 26 \pm 13$~pc\,GHz. 

The magnetic field in the core region $B_\mathrm{core} = B_\mathrm{1\,pc} r_\mathrm{core}^{-1}$, where $r_\mathrm{core}$ is the de-projected distance from the central engine to the core and equals to $r_\mathrm{core}(\nu) = \Omega_{r\nu} \nu^{-1}/\sin\theta$ \citep{1998A&A...330...79L}. The magnetic field $B_\mathrm{1\,pc}$ in Gauss at 1~pc distance can be calculated using the following equation \citep{2005ApJ...619...73H,2009MNRAS.400...26O}:
\begin{equation}
    B_\mathrm{1\,pc} \approx 0.025 \left( \frac{\Omega_{r\nu}^3 (1+z)^2}{\delta^2 \phi \sin^3 \theta } \right)^{0.25},
	\label{eq:core_shift_MF_eq}
\end{equation}
where $\phi$ is the jet half opening angle and $\phi = \phi_\mathrm{obs}\sin\theta$. We fitted the intensity along transverse jet slices measured down the ridge line with a Gaussian profile and then deconvolved it with the beam following \citet{2017MNRAS.468.4992P,2020MNRAS.495.3576K}. As a result, we measured the observed opening angle of the jet $\phi_\mathrm{obs} = \left (7\pm 1\right )^\circ$. We used the identical parameters from the approximation $\Gamma \sin \theta = 1$. Eventually, the magnetic field at 1~pc distance from the central engine $B_\mathrm{1\,pc} \approx 1.1 \pm 0.5$ G was consistent with values presented in the samples for other quasars \citep[e.g.][]{2014Natur.510..126Z}.  The magnetic field at the core region for 22.2~GHz is $B_\mathrm{core} \approx 0.4 \pm 0.2$ G. It is hard to estimate the errors of $\delta$ and $\theta$ since we only know their imprecise values. Studies of jet kinematics might help us achieve more accurate estimates later. 

As a result, we obtained similar values within the error limits with both independent approaches, the core shift method and the method using standing shock wave jump conditions and magnetic flux conservation. The fact that two different approaches are consistent with each other provides additional evidence for the shock wave assumption.

\section{Summary}
\label{sec:summary}

The distant quasar 0858$-$279 has drawn our attention due to a strong long-term variability of its peaked radio spectrum measured by RATAN-600, whereas the archival snapshot VLBI images revealed a peculiar, heavily resolved blob.
In this paper, we studied its parsec-scale structure with VLBA at multiple frequencies between 1.4~GHz and 22~GHz. We reconstructed total intensity and linear polarization maps as well as spectral index and Faraday rotation measure images. %
These data revealed a core-jet structure, as expected for a typical quasar. However, the jet was strongly curved, the core was weak and a dominant parsec-scale feature appeared at the jet bend.
The emission of the bright feature changes from the optically thin to the optically thick regime at the observed centimetre wavelengths.
We suggested that the jet bends, and a strong shock wave is formed at that region due to interaction with a dense surrounding medium.

The magnetic field inside the bright jet feature was assessed as $B = (0.55 \pm 0.37)\delta$~G within the homogeneous synchrotron source model, where $\delta$ is the Doppler factor. Taking the variability Doppler factor of 2 estimated from the RATAN-600 monitoring data, we estimated the magnetic field strength about 1~G. 
Note that this value is significantly larger than typically expected at the projected distance of 20 parsecs from the core.
It can be caused by the amplification of the magnetic field by the occurrence of a shock wave.
%
Applying the flux conservation, we calculated the magnetic field strength at the core region, and we independently estimated the same value from the core shift approach. The results were consistent, delivering a value of about 0.3~G.

We obtained high RM values from this feature above 6000~rad/m$^2$ in the observer's frame and restored the direction of the magnetic field in the jet. It is perpendicular to the jet propagation direction. Thus, all the results obtained seem to indicate the presence of a standing shock wave, where the electrons are re-accelerated.

In a follow-up VLBA study, we plan to determine the jet kinematics and investigate the variability of the dominant jet feature. 
This will allow us to check the standing shock wave hypothesis, address the scattering properties, and study this peculiar quasar in more detail.

\section*{Acknowledgements}

We thank the anonymous referee and Alan Marscher, Ilya Pashchenko, Alexander Pushkarev, Alexander Plavin and Eduardo Ros for discussions, comprehensive comments and suggestions that helped us improve this paper. We thank E.~Bazanova for the English language editing of this manuscript.
This research was supported by the Russian Science Foundation grant  16-12-10481. 
MP acknowledges support by the Spanish Ministry of Science through Grants PID2019-105510GB-C31, PID2019-107427GB-C33 and from the Generalitat Valenciana through grant PROMETEU/2019/071.
The observations with the RATAN-600 telescope of the Special Astrophysical Observatory are supported by the Ministry of Science and Higher Education of the Russian Federation. 
The VLBA is an instrument of the National Radio Astronomy Observatory, a facility of the National Science Foundation operated under the cooperative agreement by Associated Universities, Inc. 
This work made use of the Swinburne University of Technology software correlator, developed as part of the Australian Major National Research Facilities Programme and operated under licence \citep{2011PASP..123..275D}.
This research made use of the data from the MOJAVE database, which is maintained by the MOJAVE team \citep{2018ApJS..234...12L} and the University of Michigan Radio Astronomy Observatory. 
This research made use of NASA’s Astrophysics Data System Bibliographic Services.


\section*{Data Availability}

The correlated data underlying this article are available from the public NRAO archive, project code BK128.
The fully calibrated visibility and the image FITS files for the target source are made available online in CDS\footnote{\url{https://cds.u-strasbg.fr/}}.



\bibliographystyle{mnras}
\bibliography{0858}

\bsp	
\label{lastpage}
\end{document}